\documentclass{JCIN18}
\usepackage{bm}
\usepackage{cutwin}
\usepackage{graphicx}
\usepackage{subfig}
\usepackage{picins}
\usepackage[hyphens]{url}

\begin{document}

\ArticleType{Review paper}
\Year{2021}

\title{\Huge What is Semantic Communication? \\ \Large A View on Conveying Meaning in the Era of Machine Intelligence}

\author[1]{Qiao~Lan}
\author[1]{Dingzhu~Wen}
\author[1]{Zezhong~Zhang}
\author[1]{Qunsong~Zeng}
\author[1]{Xu~Chen}
\author[2]{Petar~Popovski}
\author[1]{Kaibin~Huang}

\address[1]{Department of Electrical and Electronic Engineering, The University of Hong Kong}
\address[2]{Department of Electronic Systems, Aalborg University}

\abstract{In 1940s, Claude Shannon developed the information theory focusing on quantifying the maximum data rate that can be supported by a communication channel. Guided by this fundamental work, the main theme of wireless system design up until the \emph{fifth generation} (5G) was the data rate maximization. In Shannon's theory,  the semantic aspect and meaning of messages were treated as largely irrelevant to communication. The classic theory started to reveal its limitations in the modern era of machine intelligence, consisting of the synergy between  \emph{Internet-of-Things} (IoT) and \emph{artificial intelligence} (AI). By broadening the scope of the classic communication-theoretic framework, in this article we present a view of \emph{semantic communication} (SemCom) and conveying meaning through the communication systems. We address three communication modalities: 
\emph{human-to-human} (H2H), \emph{human-to-machine} (H2M), and \emph{machine-to-machine} (M2M) communications. The latter two represent the paradigm shift in communication and computing, and define the main theme of this article. H2M SemCom refers to \emph{semantic techniques} for conveying meanings understandable not only by humans but also by  machines so that they can have interaction and ``dialogue''. On the other hand, M2M SemCom refers to \emph{effectiveness  techniques} for efficiently connecting  multiple machines such that they can  effectively execute a specific  computation  task in a wireless network. The first part of this article focuses on introducing the SemCom principles including encoding, layered system architecture, and two design approaches: (1) layer-coupling design; and (2) end-to-end design using a neural network. The second part focuses on discussion of specific techniques for  different application areas of H2M SemCom (including  human and AI symbiosis, recommendation, human sensing and care, and \emph{virtual reality} (VR)/\emph{augmented reality} (AR)) and M2M SemCom (including  distributed learning, split inference, distributed consensus, and machine-vision cameras). Finally, we discuss  the approach for designing  SemCom systems based on knowledge graphs. We believe that this  comprehensive  introduction will provide a useful guide into the emerging   area of  SemCom that is expected to play an important role in \emph{sixth generation} (6G) featuring connected  intelligence and integrated  sensing, computing, communication, and control. 
}


\maketitle


\section{Introduction} 

Modern wireless communication systems have reshaped the operations of the society and people's lifestyle, becoming an engine for propelling the data economy. Many advances in wireless systems are based on the ideas rooted at Claude Shannon's \emph{locus classicus} on information theory~\cite{Shannon1948}. In his work, Shannon defined a communication problem as one concerning ``\emph{reproducing at one point either exactly or approximately a message selected at another point}''. He  argued therein that ``\emph{semantic aspects of communication should be considered as irrelevant to the engineering problem}''. Guided by Shannon's approach and philosophy, most existing communication systems have been designed based on rate-centric metrics such as throughput, spectrum/energy efficiency, and, with the advent of 5G, latency. Nevertheless, there is an increasing belief in the community that the classic Shannon's framework needs to be upgraded for the next evolution step in communications.
Its narrow focus on the reliability level of communication starts to show its limitations in meeting the ambitious goals set for the \emph{sixth generation} (6G). In particular, the ignored \emph{meaning} behind the transmitted data are expected to play an important role in 6G communications, which places an unprecedented emphasis on machine intelligence and its interface with human intelligence. In existing systems there is a limited coupling of the high-level meaning or relevance of the data content with the transmission strategies; an example is packet prioritization based on data content, implemented in the upper networking and application layers~\cite{ICN2012,3GPP_SBA,Brown2017}. However, the separation of transmission and data's meanings and effectiveness for achieving specific goals inevitably result in redundancy, e.g., transmitting information lacking relevance or freshness. This causes the existing techniques for information filtering, transmitting, and processing to struggle with keeping pace with the exponential growth rate of data traffic~\cite{6GNetwork2021,Samsung20206GVision, B5GAI2020}. The need of highly efficient communication for supporting machine-intelligence services has triggered a paradigm shift from  ``semantic neutrality'' towards \emph{semantic communication} (SemCom)~\cite{JBao2001,JACM2012,6GNetwork2021,Shi2020FromSC}. This is the main theme of this article.

The concept of SemCom was introduced by Warren Weaver, a collaborator of Shannon, who defined a communication framework featuring three levels~\cite{Weaver1949}. The first-level, which is targeted by Shannon's information theory, aims at answering the \emph{technical problem} that ``\emph{How accurately can the symbols of communication be transmitted?}''. SemCom belongs to the second level concerning an answer to the \emph{semantic problem} that ``\emph{How precisely do the transmitted symbols convey the desired meaning?}'', beyond which the third level is defined as the \emph{effectiveness problem} that ``\emph{How effectively does the received meaning affect conduct in the desired way?}''. In Weaver's time, communication activities dominantly served the purpose of information exchange among humans. Thus the Weaver's SemCom definition should be interpreted as a concept of \emph{Human-to-Human} (H2H) communication. In the modern era of \emph{machine intelligence}, the connotation and scope of SemCom, however, have been substantially enriched and broadened to cover all three levels. This necessitates the presentation of a modern view of SemCom. 

\subsection{The Rise of Machine Intelligence}
The recent rapid advancements in \emph{Artificial Intelligence} (AI) and \emph{Internet-of-Things} (IoT) are two main factors contributing to the rise of machine intelligence. 

Research on AI dates back to 1950's. The term AI first appeared in a research proposal aimed at creating ``\emph{the embryo of an electronic computer that will be able to walk, talk, see, write, reproduce itself and become conscious of its own existence}''~\cite{Navy1958}. To materialize the vision, researchers invented neural networks to mimic the mechanism of brain neurons for processing information and realizing intelligence. Early attempts attained some success in demonstrating the effectiveness of such models, e.g. Frank Rosenblatt's famous concept of Perceptron. The single-layer linear classifier he used is widely regarded as the distant ancestor of modern \emph{Machine Learning} (ML) algorithms. The ensuing evolution of AI had experienced periodic bouts of enthusiasm interspersed with ``AI winters''. In a ``winter'', research could stay stagnant for a decade due to limited computing power, insufficient training data, and crudity of AI algorithms. These obstacles were finally eliminated in the past few years after the preceding decades of fast development of chips with remarkable number-crunching power and growing abundance of datasets. Nowadays, we witness the wide-spread use of powerful large-scale neural-network models featuring  billions to tens of billions of parameters organized in hundreds of hierarchical layers, termed the \emph{Deep Learning} (DL) architecture. Advanced ML algorithms have been designed for various tasks, including \emph{supervised learning}, \emph{unsupervised learning}, and \emph{reinforcement learning}. Via heavy-duty statistical analysis of big data, the ML algorithms can enable deep neural networks to understand the inherent patterns of physical objects and attain a wide-range of human-like capabilities, from recognition to translation.

Another paradigm shift in computing is to embed computers into tens of billions of edge devices (e.g., sensors and wearables) and connect them to the mobile networks~\cite{EconomistIoT2019,IoTsurvey2020}. Thereby, the resultant  IoT can serve as a large-scale sensor network as well as a massive platform for edge-computing. Complex tasks can be executed on the platform to improve the efficiency of businesses and the convenience of consumers. For example, sensors and cameras connected to IoT can act as a surveillance network, or save energy by smart lighting; IoT connected cows can enable the cloud to track their health conditions and eating habits, which provides useful data for smart agriculture. Individual gains may not be walloping but they are compounded as the scale of IoT grows. 

The developments of AI and IoT are intertwined and their full potential can be unleashed by integration. On one hand, AI endows on edge devices the human-like capabilities of decision making, reasoning, and vision as well as boosting their communication efficiencies and reliability. On the other hand, massive data are being continuously generated by the enormous number of edge devices in IoT (e.g., more than a hundred trillion gigabytes of data in the next 5 years~\cite{datagrowth2020}). Such data are fuel for AI and can be distilled into intelligence to support a wide-range of emerging applications and improve the efficiencies of data-driven businesses \cite{B5GAI2020}.

\subsection{The Context for Defining Semantic Communication}

The breathtaking advancements in machine intelligence and the exponential growth of machine population usher in the new era of machine intelligence. The extensive involvement of machines in modern communication give rise to three basic types of communication context: H2H communications, \emph{human-to-machine} (H2M) communications, and \emph{machine-to-machine} (M2M) communications. The classic H2H SemCom as considered by Weaver is therefore insufficient for describing future diverse communication tasks. This motivates us to broaden the scope of SemCom by defining three sub-areas matching the mentioned contexts as follows. 
\begin{itemize}
  \item \textbf{H2H SemCom:}
  The definition of H2H SemCom is consistent with the second level of the Weaver's framework and addresses the semantic problem described earlier. To be precise, the communication purpose is to accurately deliver meanings over a channel for message exchange between two human beings. To this end, the system performance is measured by how well the intended meaning of the sender can be interpreted by the receiver. 
  
  \item \textbf{H2M SemCom:} This area concerns communication between a human being and a machine. The distinction of H2M SemCom lies in the interface between human and machine intelligence, which is  different in nature and involves both the second and third level of the Weaver's framework. For H2M SemCom to be effective, the transmitted messages have to be understood not only by humans but also by machines. To be more specific, the success in H2M SemCom hinges on two aspects: 1) a message sent by a human being should be correctly interpreted by a machine so as to trigger the desired actions or responses (the effectiveness problem); 2) a message sent in the reverse direction should be meaningful for the human at the receiving end (the semantic problem). The typical applications include  human and AI symbiosis system, recommendation system, human sensing and care system, and \emph{virtual reality} (VR)/\emph{augmented reality} (AR) system. 
  
  \item \textbf{M2M SemCom:} In the absence of human involvement, M2M SemCom concerns the connection and coordination of multiple machines to carry out a computing task. Therefore, this area relates less the level-two communication (i.e., semantic problem) but more to the level-three communication  (i.e., effectiveness problem). Latest  research on M2M SemCom advocates the approach of \emph{integrated communication and computing} ($\text{IC}^2$) that promises more efficient system design under constraints on radio and computing resources. The resultant cross-disciplinary research has led to the emergence of a new class of M2M SemCom techniques to be introduced in the sequel. The typical applications include M2M SemCom are mainly related to those in the areas of distributed sensing, distributed learning, and distributed consensus (e.g., vehicle platooning). 
  
\end{itemize}

\subsection{Motivation and Outline}
SemCom has been regarded as a key enabling technology for future networks. Research on SemCom concerns the representation of semantic information, SemCom modeling, enabling techniques, and network design. A number of surveys of the area have appeared  where different  SemCom frameworks are proposed.  First, system-level issues of SemCom such as network architectures and modeling are discussed in~\cite{Popovski2020JIIS,Shi2020FromSC}. Specifically, a \emph{semantic-effectiveness} (SE) plane whose  functionalities address  both the semantic and effectiveness problems is proposed  in~\cite{Popovski2020JIIS} to realize information filtering and direct control of all layers. The new layered architecture  is  showcased with particular  applications including immersive and tactile scenarios,~\emph{integrated communication and sensing} (ISAC), and physical-layer computing. The survey in~\cite{Shi2020FromSC} focuses on SemCom modeling and semantic-aware network architecture.  Two SemCom models are introduced based on shared {\emph{knowledge graph}} (KG) and semantic entropy, respectively, each which comprises semantic encoder/decoder and semantic noise. Building on these  models, semantic networking for federated edge intelligence is then proposed to support  semantic information detection and processing, knowledge modeling, and knowledge coordination. 
On the other hand, efforts have been made to explore  SemCom enabling techniques including information representation, data transmission and reconstruction~\cite{6GNetwork2021,SemEmpower2021,Elif2021Semantic}. In particular, the work in \cite{6GNetwork2021} features  the integration of semantic and goal-oriented aspects for  6G networks and  KG based techniques for information representation, semantic information exchange measurement, semantic noise, and feedback. This survey also presents the interplay between machine learning and SemCom, identifying their mutual enhancement and cooperation in communication networks. In~\cite{SemEmpower2021}, a semantic-aware communication system is discussed from the perspective of data generation/active sampling, information transmission, and signal reconstruction. To redesign communication networks for SemCom,  conventional approaches should be revamped to support new metrics and operations such as   semantic metrics, goal-oriented sampling, semantic-aware data generation, compression,  and transmission as suggested in  \cite{Elif2021Semantic}. In view of prior work, existing SemCom frameworks are basically extended from the Weaver's classical definitions and do not comprehensively incorporate current advancements of relevant technologies. There is still a lack of a systematic survey article that provides a unified framework of SemCom in the era of machine intelligence; this is precisely the motivation for the current work.

The contributions of this paper can be summarized as follows. First, this paper defines three different areas of SemCom, i.e. H2H SemCom, H2M SemCom, and M2M SemCom, by identifying the involved subjects and objects. The proposed framework can accordingly describe existing technologies, models, and frameworks, providing a comprehensive reference for both researchers and practitioners. Next, with the proposed framework, we conclude current advancements of technologies that are relevant to or beneficial for SemCom, which can help readers in interpreting easier their research in the context of SemCom. Furthermore, we incorporate the  {KG based SemCom} technologies and extend their applications into H2M SemCom, H2M SemCom, and M2M SemCom scenarios. In addition, according to existing 6G visions, potential technologies and use cases that are helpful for SemCom are introduced.

While H2H SemCom is a classic, well studied area, our discussion focuses on H2M and M2M SemCom for their being new paradigms in the modern era of machine intelligence. Furthermore, we propose a new direction of KG based SemCom that helps accomplish H2H SemCom, H2M SemCom, and M2M SemCom by exploiting the semantic representations of information.  An overview of SemCom techniques and applications  covered in this article is provided in in TABLE~\ref{tbl:list}.

\begin{table*}[t]
   \caption{Summary of Advancements in SemCom Techniques and Applications}
    \label{tbl:list}
    \vspace*{-5pt}
    \small
    \centering
    \begin{tabular}{ p{2.5cm} p{12.5cm}}
        \hline
        \hline
        \textbf{SemCom Areas} & \textbf{Advancements}   \\
        \hline
            H2M SemCom &  \vspace*{-8pt}
                            \begin{itemize}
                            \item \emph{Human-machine symbiosis:} AI-assisted systems~\cite{xu2017new,ranoliya2017chatbot,lysaght2019ai,machado2018ai,svyatkovskiy2019pythia, ziebinski2017review, kannan2018new}, interactive ML~\cite{holzinger2013human,holzinger2016interactive,berg2019ilastik, schneider2020human, de2013exploring}, worker-AI collaboration~\cite{sriraman2017worker, AMT, okamura2009haptic, sornkarn2016can, chakraborti2019can}.
                            \item \emph{Recommendation}: Emotional health monitoring \cite{rosa2018knowledge}, tourist \cite{gavalas2011web}, entertainment~\cite{xia2015reciprocal, zhao2018analyzing, chin2010cprs, davidson2010youtube}, remote healthcare \cite{zhang2015cadre, wang2016design}.
                            \item Human sensing and care: Home monitoring \cite{suryadevara2012wireless,lin2006wireless, kelly2013towards}, super soldier~\cite{windau2013situation,golestani2020human}, human activity recognition~\cite{windau2013situation,golestani2020human}, remote healthcare~\cite{liszka2004keeping,WP1,WP2}.
                            \item \emph{VR/AR}: Techniques  \cite{VRARRen2019,elbamby2018toward,chen2018virtual,yu2019skin,HuaweiXRlatency2018} and applications \cite{VRARsurvey1,VRARsurvey2}.
                            \item \emph{Latent semantic analysis (LSA)}  \cite{LSA2}.
                            \item \emph{Computation offloading for  edge computing}~\cite{wang2020edge,tang2018novel}.
                            \item \emph{Decentralization for privacy  preservation} ~\cite{corchado1995distributed,armknecht2011efficient}.
                            \end{itemize} 
                            
                            \vspace*{-15pt}
                            \\
        \hline
            M2M SemCom & \vspace*{-8pt}
                            \begin{itemize}
                            \item \emph{Distributed learning}: Local gradient computation~\cite{Deniz2020TSP,Deniz2020TWC,Tassiulas2019Arxiv,Nader2021Infocom,ZZZ2021Arxiv}, over-the-air computing~\cite{GX2020TWCBAA,GXAirCompMag,Liu2020CL,Shi2020TWC}, importance-aware RRM~\cite{Dongzhu2021TWC, Ren2020TWC,Tao2021TWC, HKB2021TCCN}, differential privacy~\cite{Seif2020ISIT,DZ2021JSAC}.
                            \item \emph{Split inference}: Feature extraction~\cite{Zhang2020CM,Bajic2021TIP,Bajic2021TIP, Ko2018AVSS, Pagliari2020TC, JunZhang2021ICASSP}, importance-aware RRM ~\cite{QiaoLan2021TechReport,Zhuang2018NIPS,ShengZhou2019Infocom,Bajic2021TIP, Park2021Sensors,Choi2018CVPR}, SplitNet approach~\cite{Gunduz2021JSAC,Lee2019Access,XuChen2020TWC,Bennis2021arxiv}.
                            \item \emph{Distributed consensus}: Local-sate estimation and prediction~\cite{Winnie2016TITS,Monica2017TITS,Siddique2019IJDSN}, SDT~\cite{XUE2020AutoCons}, PBFT consensus~\cite{Li2021TPDS}, Vehicle platooning~\cite{Hult2016spm,JPark2021arxiv, Niu2020WC}, Blockchain~\cite{Petar2019IoT-J,XUE2020AutoCons}.
                            \item \emph{Machine-vision cameras}: ROI-based effectiveness encoding~\cite{Sultana2019Access,Ren2018Network}, camera-side feature extraction~\cite{Kot2020TIP}, production-line  inspection~\cite{Ota2018TII}, surveillance~\cite{Chen2017TCSVT,Sultana2019Access}, aerial and space sensing~\cite{Skinnemoen2014AERST}.
                            \end{itemize} 
                            \vspace*{-15pt}
                            
                            \\ 
        \hline
            KG based SemCom & \vspace*{-8pt}
                            \begin{itemize}
                            \item \emph{Enhancement for AI applications}: FAQs~\cite{questionAnswering1,questionAnswering2}, virtual assistants~\cite{virtualAssistants1,virtualAssistants2}, dialogue~\cite{Dialogue}, recommendation~\cite{Recommendation1,Recommendation2,Recommendation3}.
                            \item \emph{KG based H2H SemCom}: Semantic coding~\cite{ERNIE1,ERNIE2,SEED,knowledgeSurvey,ErrorCorrection1,ErrorCorrection2}.
                            \item \emph{KG based  H2M SemCom}: \cite{humanMachineRequirement,ContextKnowledge,Recommendation3,Dialogue,Imitation1,Imitation2}. 
                            \item \emph{KG based M2M SemCom}: KG construction and update~\cite{Distributed1, Distributed2,Fusion1,Fusion2,Fusion3,Fusion4}, KG based network management~\cite{Manage1,Manage2,Manage3}, interpretation for cross-domain applications~\cite{M2M1,M2M2,SensorWeb}.
                            \end{itemize} 
                            \vspace*{-15pt}
                            \\
        \hline
    \end{tabular}
\end{table*}

The remainder of the paper is organized as follows. In Section~\ref{Section:principles}, we introduce the SemCom principles including semantic and effectiveness encoding, a new network layered architecture, and design approaches. Next, semantic/effectiveness encoding and transmission techniques targetting specific application areas of  H2M SemCom and M2M SemCom are presented  in the following  two sections. Specifically, in Section~\ref{Section:H2MSemCom}, semantic encoding and H2M SemCom techniques are discussed for the areas of human-machine symbiosis, recommendation, human sensing and care, and VR/AR. Section~\ref{Section:M2MSemCom} focuses on M2M SemCom including effectiveness encoding and SemCom techniques for the areas of  distributed learning, split inference, distributed consensus and machine-version cameras. Subsequently, we introduce the KG based SemCom approach in Section~\ref{Section:KG}. Finally, in Section~\ref{Section:conclusion}, a visino of SemCom in the 6G era is proposed. 

For ease of reference, We summarize the definitions of the acronyms that are used in this paper in Table~\ref{tbl:acronym}.

\begin{table*}[t]
    \caption{Summary of Acronyms}
    \label{tbl:acronym}
    \vspace*{-5pt}
    \small
    \centering
    \begin{tabular}{ p{1.2cm} p{6.2cm} p{1.2cm} p{6.2cm}}
        \hline
        \hline
        \textbf{Acronym}      & \textbf{Definition}                                     & \textbf{Acronym}  & \textbf{Definition}                       \\
        \hline
        6G                         & Sixth Generation                                              & LSTM     & Long-Short-Term Memory                          \\
AE                         & Auto-Encoder                                                  & M2M      & Machine-to-Machine                              \\
AI                         & Artificial Intelligence                                       & MIMO     & Multiple Input-Multiple Output                  \\
AirComp                    & Over-the-Air Computing                                        & ML       & Machine Learning                                \\
AR                         & Augmented Reality                                             & MLP      & Multi-layer  Perceptron                         \\
BERT                       & Bidirectional   Encoder Representations from Transformers     & mMTC     & massive Machine Type Communication              \\
CDD                        & Channel Decoded Data                                          & MR       & Mixed Reality                                   \\
CNN                        & Convolutional Neural Network                                  & MSE      & Mean  Squared  Error                            \\
CRI                        & Channel Rate Information                                      & NLP      & Natural Language Processing                     \\
CSI                        & Channel  State  Information                                   & PAI      & Partial Algorithm Information                   \\
DII                        & Data Importance Information                                   & PBFT     & Practical Byzantine Fault Tolerance             \\
DL                         & Deep Learning                                                 & PCA      & Principal Component Analysis                    \\
DNN                        & Deep Neural Network                                           & QoS      & Quality-of-Service                              \\
DTI                        & Data Type Information                                         & RNN      & Recurrent Neural Network                        \\
ECG                        & Electrocardiogram                                             & ROIs     & Regions of Interests                            \\
eMBB                       & enhanced   Mobile Broadband                                   & RRM      & Radio-Resource  Management                      \\
EMG                        & Electrocardiogram                                             & SE       & Semantic-Effectiveness                          \\
FEEL                       & Federated Edge Learning                                       & SEED     & Semantic/Effectiveness Encoded Data             \\
FL                         & Federated Learning                                            & SemCom   & Semantic Communication                          \\
FoV                        & Field-of-View                                                 & SGD      & Stochastic Gradient Descent                     \\
H2H                        & Human-to-Human                                                & SMCV     & Squared Multi-Variate Coefficients of Variation \\
H2M                        & Human-to-Machine                                              & SplitNet & Split  Neural  Network                          \\
IB                         & Information   Bottleneck                                      & SVD      & Singular Value Decomposition                    \\
IC$^2$                     & Integrated Communication and Computing                        & UAV      & Unmanned Aerial Vehicle                         \\
IoE                        & Internet-of-Everything                 & URLLC    & Ultra-Reliable Low-Latency Communication        \\
IoT                        & Internet-of-Things                                            & VR       & Virtual Reality                                 \\
ISAC                       & Integrated Communication and Sensing                        & XR       & Extended Reality                                \\
KG                         & Knowledge Graph                                               & SDT      & Semantic Difference Transaction                 \\
LNA                        & Linear Analog Modulation                                      & MLM      & Masked Language Model                           \\
LSA                        & Latent Semantic Analysis                                      &          &         \\
        \hline                           
    \end{tabular}
\end{table*}

\section{SemCom Principles: Encoding, Architecture, and Design Approaches}\label{Section:principles}

\begin{figure*}[t]
    \centering
    \subfloat[Communication System in Shannon's Theory]{\label{Fig: ShannonCom}
    \includegraphics[width=0.80\textwidth]{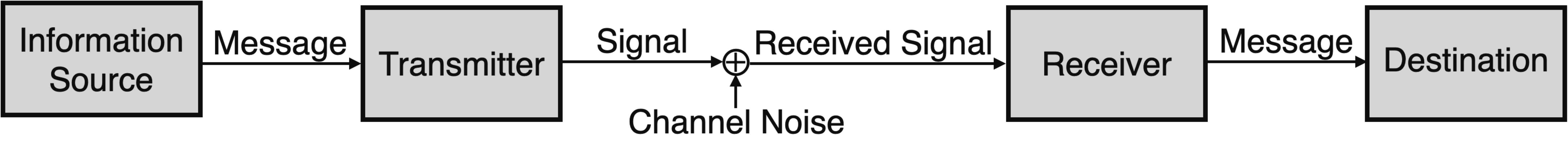}}\vspace{4mm}
    \subfloat[Semantic Communication System]{\label{Fig: SemCom}
    \includegraphics[width=0.75\textwidth]{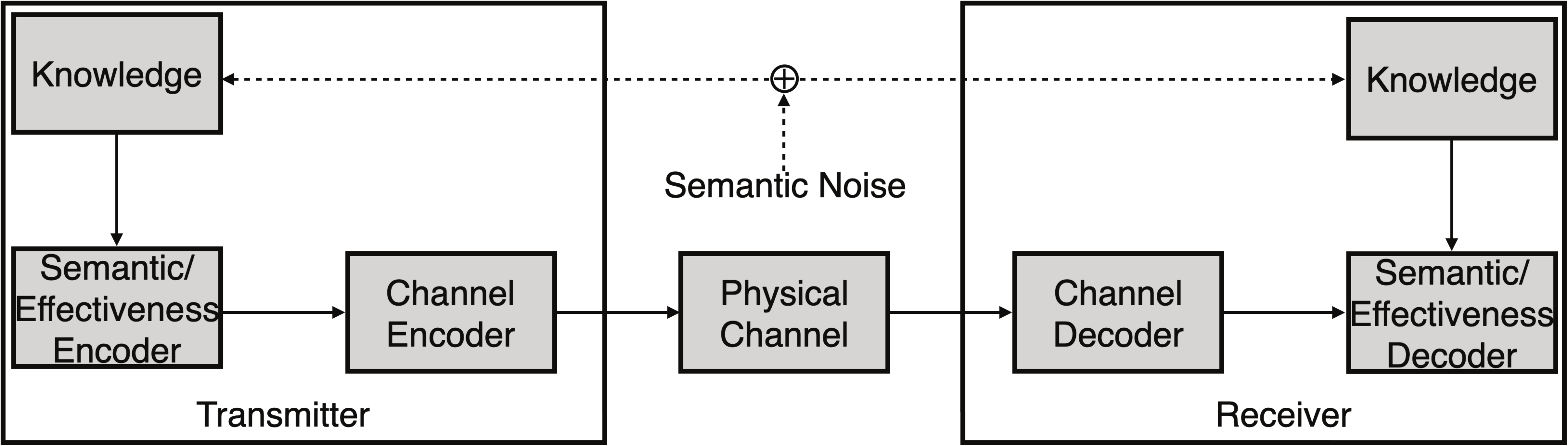}}
    \caption{Models for a point-to-point communication a) Shannon's model and b) Model of a SemCom system.}
\end{figure*}

\subsection{Encoding for SemCom}\label{sec:encodingforSemCom}

In~\cite{Shannon1948}, the fundamental problem of communication was described as that of reproducing at one point either exactly or approximately a message selected at another point. The communication-theoretic model of Shannon consists of five parts as illustrated in Fig. \ref{Fig: ShannonCom} and explained as follows.  
\begin{enumerate}
    \item An \emph{information source} produces messages to be transmitted to the receiver. 
    \item A \emph{transmitter} encodes and modulates the messages  into  a signal for robust transmission over a  unreliable  channel.
    \item A  \emph{channel} is the medium used to propagate  the encoded  signal from the transmitter to receiver. In the propagation process, the external, random disturbance to the signal is  called \emph{channel noise}. 
    \item A \emph{receiver} performs decoding and demodulation to reconstruct the transmitted message from the received signal such that errors due to channel distortion  are corrected; 
    \item A \emph{destination} is a  human being  or a  machine for whom the message is intended. 
\end{enumerate}
Information-theoretic encoding focuses on the statistical properties of messages instead of the content of messages. The transmitted  message is one selected from a set of possible messages with a given distribution. Mathematically, information theory simplifies H2H communication to  transmission of a finite set of symbols. Nevertheless, in practice, the messages have \emph{meaning, relevance, and/or usefulness}. To be specific, they refer to or are correlated with certain physical or conceptual entities or are contributing towards the achievement of some goal. This  semantic aspect of communication was  originally treated in Shannon's theory as being irrelevant to the engineering problem of information transmission. For example, a tacit assumption in Shannon's model is that the sender always knows what is relevant for the receiver and the receiver is always interested and ready to receive the data sent by the transmitter.

As mentioned earlier, Weaver proposed a more general communication framework characterized by   three levels of problems, namely the technical problem (solved by Shannon's theory), semantic problem, and effectiveness problem   \cite{Weaver1949}.  While Weaver's framework targets H2H communication, we consider the modern SemCom in the era of machine intelligence  as addressing both the semantic and effectiveness problems. A diagram of the SemCom system is presented in Fig. \ref{Fig: SemCom}. Accordingly, there are two classes of techniques: 
\begin{itemize}
\item \emph{Semantic encoding and transmission:} This class of techniques target scenarios where the destination is a  \emph{human being} (e.g., H2H and M2H SemCom).  Their purpose is to  convey the meaning of a  transmitted message as accurate as possible so that it can be correctly interpreted by a human. Therefore, the design of such techniques is to solve  the semantic  problem in Weaver's framework.

\item \emph{Effectiveness encoding and transmission:}  This class of techniques target scenarios where the destination is a \emph{machine}   (e.g., H2M and M2M). Then  the techniques aim at delivering a message as a instruction or query  to the machine such that it can perform what the sender requires it to do or respond appropriately. In this sense, their design focuses on the effectiveness aspect of communication, thereby the name. 
\end{itemize}

In the remainder of this sub-section, the principles of  semantic and effectiveness encoding are introduced while application specific  techniques  are discussed in following sections. 

\subsubsection{Semantic Encoding}
Even though Shannon's theory does not explicitly target SemCom, information-theoretic encoding can be adopted for the latter  by extending  two key notions, \emph{entropy} and \emph{mutual information}, to define \emph{semantic entropy} and \emph{semantic mutual information}. The entropy of a discrete source measures the amount of information in each sample and depends on the source's statistics. Mathematically, the entropy of a message $X$ is defined as $H(X)=-\sum_{i=1}^np(x_i)\log{p(x_i)}$, where $x_1,x_2,\cdots,x_n$ are possible outcomes of $X$ with probabilities $p(x_1),p(x_2),\cdots,p(x_n)$. Accordingly, the mutual information is given by $I(X;Y)=H(X)-H(X|Y)$, which indicates how much the amount of information about the transmitted message $X$ is obtained after receiving the message $Y$. On the other hand, the mutual information between the source and destination quantifies the amount of information obtained about the former by observing the latter. The combined use of the two measures allows the study of the maximum data rate under a constraint of ``physical distortion" [e.g., \emph{mean squared error} (MSE)]. The unsuitability of these information-theoretic  measures for SemCom due to their lack of semantic elements is obvious by considering the following example.  A single-letter error results in a transmitted word of ``big" to be received as ``pig"; the reception of the  word ``cattle'' due to the transmission of ``cow'' corresponds to errors in multiple letters. The former represents much more reliable information transmission  than the latter but the reverse is true from the perspective of semantic transmission. 

An attempt on defining the semantic entropy was made  in~\cite{JBao2001}. Therein, a semantic source is modeled as a tuple $(W,K,I,M)$ with $W$ modeling the observable world that includes a set of interpretations, $K$ representing source's background knowledge, $I$ indicating source inference that is relevant to background knowledge, and $M$ denoting message generator or encoder. Then given the probability $\mu(w)$ for each elements in $W$, let $W_x$ indicate the subset of $W$ in which the message $x$ is justified as ``true'' by the inference $I$, i.e., $W_x=\{w\in W|w\overset{I}{=}x\}$, and the \emph{logical} probability of $x$ is defined as $p(x)=\frac{\sum_{w\in W_x}\mu(w)}{\sum_{w\in W}\mu(w)}$ and the corresponding semantic entropy is $H_s(X)=-\sum_x p(x) \log p(x)$. Those definitions lay a foundation for semantic encoding/decoding and semantic  transmission. For instance, the recent work in \cite{Shi2020FromSC} argues that the key issue of SemCom is to find a proper semantic interpretation $W$ (also termed as semantic representation) and the coding scheme $P(X|W)$ such that the semantic ambiguity of transmitted message $H(W|X)$ and coding redundancy $H(X|W)$ are close to zero. Consequentially, the model (semantic) entropy $H(W)$ and the message (syntactic) entropy follows from $H(X) = H(W) + P(X|W) - H(W|X)$, meaning that the semantic encoder could achieve intentional source compression with an information loss $H(W)-H(X)$.

Departing  from the above theoretic abstraction, there are diversified approaches for designing practical semantic encoding. The first approach is KG based semantic-encoding that is decomposed into two stages: 1) finding a  proper representation of common knowledge background of the communication parties in the form of a KG; 2) encoding  data using the KG. Detailed discussion is presented in Section V. Second, the power of ML gives rise to the  learning based approach of integrated semantic and channel (i.e., information theoretic) encoding.  As an example, for text transmission, a joint semantic  and channel coding scheme based on deep learning is proposed in~\cite{Qin2021DeepSC}, where encoding a sentence $\mathbf{s}$ is represented by $\mathbf{x} = C_{\mathbf{\alpha}}(C_{\mathbf{\beta}}(\mathbf{s}))$ with $C_{\mathbf{\alpha}}(\cdot)$ denoting the channel encoder and $C_{\mathbf{\beta}}(\mathbf{s})$ denoting the semantic encoder. It follows that the decoding process is modeled by $\hat{\mathbf{s}} = C_{\mathbf{\chi}}^{-1}(C_{\mathbf{\delta}}^{-1}(\mathbf{y}))$ with the received signal $\mathbf{y}$. The  encoders and decoders are trained as a single neural network by treating the channel as one layer in the model (similar to SplitNet discussed in the sequel). The training process features the  consideration of both semantic similarity and transmission data rate. Specifically, the sentence similarity between the original sentence $\mathbf{s}$ and the recovered sentence $\hat{\mathbf{s}}$ is given by
\begin{equation}\label{eqn:similarity}
    \text{match}(\mathbf{s},\hat{\mathbf{s}})=\frac{\mathbf{B}(\mathbf{s})\cdot\mathbf{B}(\hat{\mathbf{s}})^{\top}}{\|\mathbf{B}(\mathbf{s})\|\|\mathbf{B}(\hat{\mathbf{s}})\|},
\end{equation}
where $\mathbf{B}(\cdot)$ represents \emph{bidirectional encoder representations from transformers} (BERT), a well-known model used for semantic information extraction~\cite{Qin2021DeepSC,LuSimilarity2021} (more details are presented in Section \ref{subsec:BERT}). Another approach is based on  \emph{latent semantic analysis} (LSA) which 
compresses  text documents by finding their low dimensional semantic representations. This is achieved  by finding a low-dimensional semantic subspace using \emph{singular value decomposition} (SVD) of document-term matrices that indicates appearances of  specific words in the documents and then projecting these matrices onto the the subspace (see more details in  Section~\ref{subsec:LSA}).

\subsubsection{Effectiveness Encoding}
\label{subsec: effectiveness_encoding}
Effectiveness encoding is to compress messages while retaining their effectiveness as instructions and commands for machines. Techniques are system and application specific and thus there exist a wide-range of designs. As examples, we discuss effectiveness source encoding targeting two  representative tasks:  classification and distributed machine learning. More application specific effectiveness encoding techniques are discussed in Section~\ref{Section:M2MSemCom}. 

\begin{itemize}
    
    \item \emph{Information Bottleneck (IB) for Classification:}
    Consider source encoding of an information source represented by a random variable  $X$.  Classic coding schemes based on  Shannon's rate-distortion theory aims at finding a representation close to $x$ in terms of MSE. On the contrary, IB is aware of the computing  task (e.g., classification), denoted as  $Y$, making it an  effectiveness coding scheme. The main feature of IB is to extract information $\tilde{X}$ from the signal source $X$ that  can contribute to the effective execution of $Y$ as much as possible. Taking classification for instance, $\tilde{X}$ shall represent the most discriminative feature of $X$. Mathematically, the IB design aims at finding the optimal tradeoff between maximizing the compression ratio  and the preserved  effectiveness  information, corresponding to simultaneously minimizing the mutual information $I(\tilde{X};X)$ and maximizing $I(\tilde{X};Y)$. This is equivalent to solving the following multi-objective optimization problem~\cite{Tishby1999AIB}: 
    \begin{equation}
        \label{eqn: IB_Design}
        \min\limits_{p(\tilde{x}|x)}I(\tilde{X};X) - \beta I(\tilde{X};Y)
    \end{equation}
    where the conditional distribution $p(\tilde{x}|x)$ denotes the source encoder and $\beta$ is a combining weight. Its  optimal solution is task-specific. For classification, $Y$ is the label predicted  by the classifier. A general algorithm constructs the optimal source encoder in~\eqref{eqn: IB_Design} via alternating iterations. In each iteration, the probability density functions $p(\tilde{x})$, $p(y|\tilde{x})$ and $p(\tilde{x}|x)$ are  determined step-by-step. The IB has attracted attention in the area of machine learning as it contributes the much needed theory for studying deep learning~\cite{Tishby2015ITW}. In particular,  training a feature-extraction encoder in a \emph{deep neural network} (DNN) can be interpreted as solving a IB-like problem where  the encoder's function is to encode  an input sample  $x$ into a compact feature map $\tilde{x}$. The encoding operations, e.g., feature compression and filter pruning, regulate the  discussed tradeoff in IB.    
    
    \item \emph{Stochastic Gradient Quantization:} One common  method of implementing \emph{federated learning} (FL), a popular distributed-learning framework based on SGD, requires a device to compute and transmit to a server a \emph{stochastic gradient}, computed by taking derivative of a loss function with respect to the parameters of a AI model under training. Detailed discussion of FL is provided in Section~\ref{sec:M2MDL}. A gradient is high dimensional as its length is equal to the number of model parameters. For instance, the popular Resnet18 model has around $11$ million trainable parameters. As it's transmission incurs excessive communication overhead, a gradient should be compressed by quantization at the device. A generic vector quantizer is unsuitable   for two reasons. \emph{First,} its design is based on using the MSE as the distortion  metric. This metric is undesirable for the current task since a gradient conveys a gradient-descending direction on a (learning) loss function and this metric fails to directly  reflect the direction deviation. \emph{Second,} the conventional vector quantizer can handle only low-dimensional vectors because its complexity grows exponentially fast with the vector dimensions. To tackle  the two challenges calls for the design of new effectiveness techniques for source encoding of stochastic gradients. One such design is presented in \cite{GrassQuantization2020} with two key features. The first is to divide a high-dimensional gradient into many low-dimensional blocks, each of which is quantized using a low-dimensional \emph{component quantizer}. The results are combined to give the high-dimensional quantized gradient with combining weights optimized to minimize descent-direction distortion. The second feature is to design a component quantization using the method of a \emph{Grassmannian manifold}. In the current design, the manifold refers to a space of lines where each line (plus the sign of an associated combining weight) suitably represents a particular descent direction. Essentially, the codebook of a  Grassmannian quantizer comprises a set of unitary vectors that are optimized to minimize the expected directional  distortion. The effectiveness source encoder for stochastic gradient, designed to target the task of FL, is shown to achieve close-to-optimal learning performance while substantially reducing  the communication overhead with respect to the state-of-art approaches, such as a binary gradient quantization scheme called signSGD \cite{GrassQuantization2020}. 
    
\end{itemize}

\subsection{SemCom  Architecture and Design Approaches} 
\label{subsec: semcom_design_approaches}
\begin{figure*}[t!]
    \centering
    \includegraphics[width=0.95\textwidth]{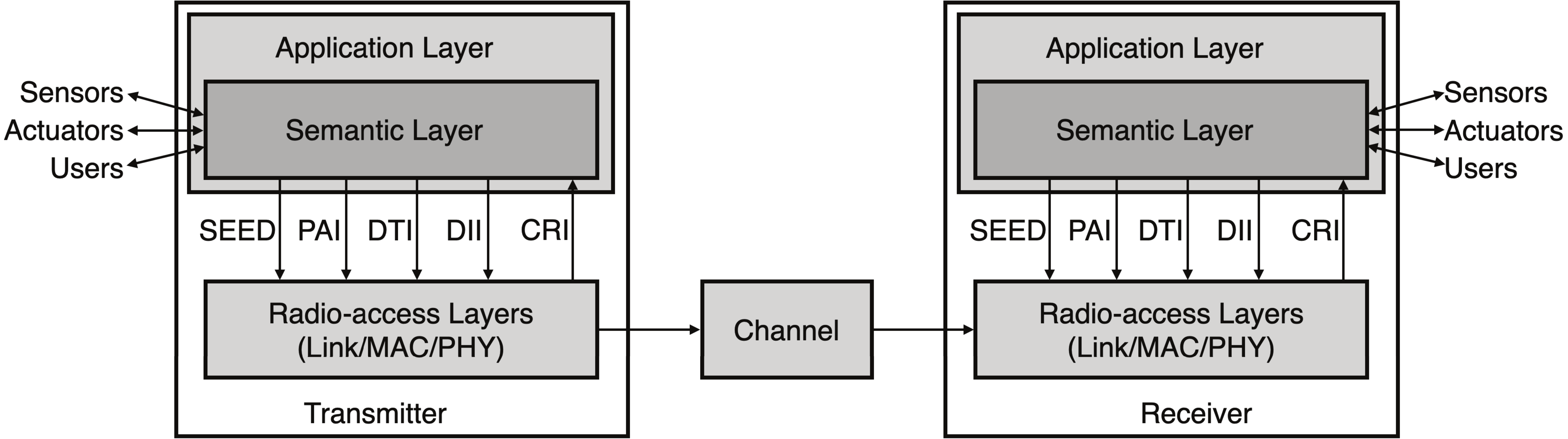}
    \caption{The layered architecture of a SemCom system and inter-layer messages.}\label{Fig: layered architecture}
\end{figure*}

The protocol stack  of a radio access network is modified in \cite{Popovski2020JIIS} to support SemCom. Its key feature is the addition of a that interacts with and control all layers to  provide efficient solutions for  both semantic and effectiveness problems in Weaver's framework. In this subsection, we propose a new, simpler SemCom architecture as  shown in Fig. \ref{Fig: layered architecture}.  It builds on the conventional protocol stack but adds a \emph{Semantic Layer} that resides in the Application Layer  as a sub-layer. This allows the Semantic Layer to
interface with sensors and actuators, have access to algorithms and content of data in the specific application. The main functions of the Semantic Layer is to perform semantic/effectiveness encoding/decoding discussed in the preceding subsections. On the other hand, the techniques for radio access layers (i.e., Physical Layer, Medium Access Layer, and Logical Link Control  Layer) are  largely  derivatives of Shannon's information theory; their design is focused on improving semantic-agnostic performance  such as  data rate, reliability, and latency. Then Semantic Layer transmits to lower layers \emph{Semantic/Effectiveness Encoded  Data} (SEED)  and receives from them \emph{Channel Decoded Data}s (CDD).  Based on the proposed architecture, we describe two approaches to the SemCom system design,  layer-coupling approach and \emph{split neural network} (SplitNet), respectively.

\subsubsection{Layer-coupling Approach}\label{sec.layers}
The first approach, called layer-coupling approach, is to jointly design the Semantic Layer  and radio access layers. To this end, we propose the possibility of exchanging control signals  between them (see Fig. \ref{Fig: layered architecture}). Among others, a set of basic   signals are defined and their functions in layer-coupling design described as  follows. 
\begin{itemize}
    \item \emph{Channel Rate Information (CRI):} The information fed back  from lower layers  enables the \emph{semantic/effectiveness} (SE)  encoders to adapt their coding rates to the wireless channel state.
    
    \item \emph{Data Importance Information (DII):} It measures the heterogeneous importance levels of elements of the SE encoded data. For a human receiver it is the interpretation, while for a machine that acts as a receiver is the is effective execution of a task. Examples include identifying keywords in a sentence in terms of representation of its semantic meaning  or discriminate gains of different features of an image for the purpose of classification. Such information facilitates adaptive transmission, multi-access, and resource allocation in the lower layers. For instance, for  data uploading to a server  for model training, the uncertainty levels of local samples can be used as DII to schedule devices \cite{HKB2021TCCN}. 
    
    \item \emph{Partial Algorithm Information (PAI)} that includes essential characteristics of current algorithms, such as information related to the AI-architecture or the target function in distributed computing (e.g., average or maximum). PAI enables the physical layer to deploy effectiveness transmission techniques, such as AirComp discussed in the preceding sub-section. 
    
    \item \emph{Data Type Information (DTI)} that indicates which category the data belongs to. This includes, for example, image for machine recognition, image for human, and stochastic gradient for machine learning, etc. DTI enables the radio access layers to choose transmission techniques  based on a suitable performance metric (e.g., Grassmannian quantization for gradient source encoding discussed in the last sub-section)  and understand the corresponding performance requirements (e.g., an image is more sensitive to noise for human vision than for pattern recognition). 
    
\end{itemize}
The above control signals can be transmitted over a control channel to the receiver and used by its semantic layer to  remove semantic noise from CDD for semantic symbol error correction or control computing at the Application Layer. The relevance of the above controls signals to techniques discussed in the sequel are summarized in Table \ref{tbl:acronym_lc}. 

\begin{table*}[t]
    \caption{Examples  of encoded data and control signals  in the SemCom architecture.}
    \label{tbl:acronym_lc}
    \vspace*{-5pt}
    \small
    \centering
    \begin{tabular}{p{3cm} p{11cm}}
        \hline
        \hline
        \textbf{Signal/Data}                   & \textbf{Examples} \\
        \hline
        SEED                                &  \vspace{-10pt}  \begin{itemize}
            \item Compressed documents by projection onto a semantic space (Section~\ref{sec:H2Msymbiosis});
            \item Local stochastic gradients (Section~\ref{sec:M2MDL});
            \item Features of data samples (Section~\ref{sec:M2MSI});
            \item Preference similarity scores (Section~\ref{sec:encodingforSemCom},~\ref{sec:H2Msymbiosis});
            \item Characteristics of  biomedical signals (Section~\ref{sec:H2Msensing}); 
            \item Field-of-view for VR/AR (Section~\ref{sec:H2MVRAR});
            \item Temporal ROIs of multimedia sensing data  (Section~\ref{sec:M2MDS}). \vspace{-10pt}
        \end{itemize}                \\
        PAI                          &      \vspace{-10pt}
        \begin{itemize}
            \item Aggregation and SGD in federated learning  (Section~\ref{sec:H2MVRAR});
            \item Feature extraction method and inference model (Section~\ref{sec:H2Msymbiosis}~\ref{sec:M2MSI}); 
            \item Distributed consensus algorithm (e.g., vehicle platooning and blockchain) (Section~\ref{sec:M2MDD}).    \vspace{-10pt}
        \end{itemize}
                    \\
        DTI                                &     \vspace{-10pt}
        \begin{itemize}
            \item Sensing data type (Section~\ref{sec:H2Msensing},~\ref{sec:H2MVRAR},~\ref{sec:M2MDS});
            \item Type of biomedical signals (Section~\ref{sec:H2Msensing});
            \item Type of local model update (Section~\ref{sec:M2MDL}).      \vspace{-10pt}  
        \end{itemize}
                    \\
        DII                           &    \vspace{-10pt}
        \begin{itemize}
            \item Raw data importance (Section~\ref{sec:H2Msensing},~\ref{sec:H2MVRAR}); \item Feature importance (Section~\ref{sec:H2Msensing},~\ref{sec:H2MVRAR},~\ref{sec:M2MSI});
            \item Gradient importance (Section~\ref{sec:M2MDL});
            \item Spatial ROIs (Section~\ref{sec:M2MDS}). \vspace{-10pt}
        \end{itemize}
        
                 \\

        \hline                           
    \end{tabular}
\end{table*}

\begin{figure*}[t!]
    \centering
    \subfloat[SemCom System based on Traditional Source-Channel Separation Coding]{\label{Fig: SSCC}
    \includegraphics[width=0.98\textwidth]{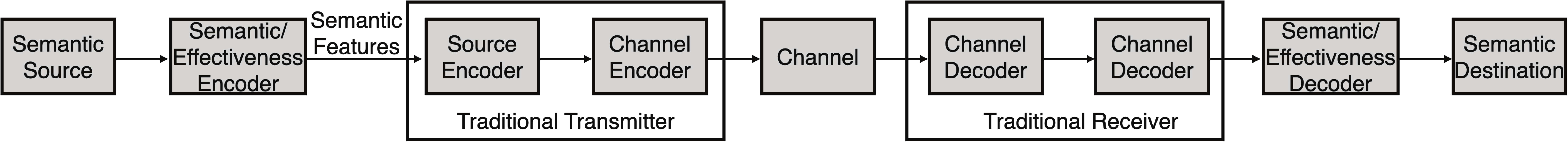}}\vspace{4mm}
    \subfloat[SemCom System Using the SplitNet Approach]{\label{Fig: JSCC}
    \includegraphics[width=0.9\textwidth]{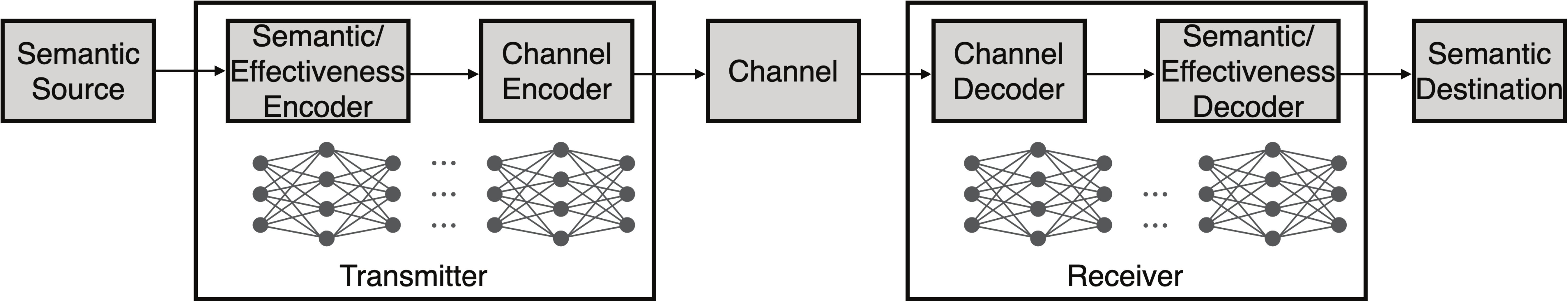}}
    \caption{Comparison between (a) the traditional source-channel separation coding, and (b) the SplitNet approach.}
\end{figure*}

\subsubsection{SplitNet Approach}
\label{sec:jscc}
An extreme form of layer-coupling design is to integrate Semantic Layer and Physical Layer into a single end-to-end global DNN \cite{Qin2021DeepSC}. The global DNN is split into two parts, namely encoder and decoder and the communication channel is sandwiched between them. This is termed SplitNet and its architecture shown in Fig.~\ref{Fig: JSCC}. The encoder model (decoder model) is further divided into two sub-modules, semantic encoder (decoder) and channel encoder (decoder),  each of which by itself is a  neural network \cite{Qin2021JSAC,Qin2021DeepSC-S}. This facilitates training in practice (see more details in Section~\ref{sec:M2MSI} about split inference). Note that the new channel encoder (decoder) in the SplitNet replaces the  source and channel encoders (decoders) in the conventional digital architecture in Fig.~\ref{Fig: SSCC}. The new encoder  directly transmits analog modulated symbols instead of quantizing them into bits and mapping them to predefined modulation symbols.

SplitNet is closely related to the area of \emph{joint source-channel coding}. The optimality of \emph{source-channel separation} was proved by  Shannon in the case of a point-to-point link with asymptotically large code blocklength \cite{Shannon1948}. This simplifies the design of communication system as source encoder/decoder and channel encoder/decoder can be optimized as separate modules. This  has become a feature of classic design approaches and led to the establishment of source and channel coding as separate sub-disciplines~\cite{cover1999elements}. However, source-channel separation is sub-optimal  in the regime of finite code length \cite{fresia2010jscc}. It worth mentioning that the sub-optimality is also shown in the context of SemCom  \cite{JBao2001}. In practice, given a finite bit-length budget (e.g., short packet transmission), the end-to-end signal distortion, or the reconstruction quality of transmitted information, sees  a complex  tradeoff between  source and channel decoding errors. This has motivated researchers to explore  the approach of joint source-channel coding with finite code lengths  \cite{choi2019, csiszar1982, kostina2013}. The joint design has been shown to be simpler and potentially more effective than its separation counterpart in practical SemCom applications, such as transmission of multimedia content \cite{Bursalioglu2013,Kurka2020,Yang2021,Bourtsoulatze2019,Katsaggelos2005,Gunduz2021JSAC}, speech \cite{Qin2021DeepSC-S}, and text \cite{Qin2021DeepSC,Farsad2018}. In particular, the notion of deep joint source-channel coding has appeared in \cite{Kurka2020,Yang2021,Bourtsoulatze2019,Gunduz2021JSAC,Qin2021DeepSC-S}, where both the source encoder (decoder) and channel encoder (decoder) are implemented by DNNs. For example, the image retrieval problem in the context of wireless transmission for remote inference is considered in \cite{Gunduz2021JSAC}. In their joint source-channel coding approach, the feature vectors are mapped to the channel symbols and decoded at the receiver, where the source and channel encoders are integrated by a DNN after the feature encoder while the the source and channel decoders are consolidated by a DNN, followed by a fully-connected classifier. 

Most recently,  SplitNet was also adopted in an end-to-end design of a SemCom system. For example, the SplitNet design presented  in \cite{Qin2021DeepSC} for SemCom system is built on the deep-learning based \emph{natural language processing} (NLP). The key component of the design uses  a  \emph{Transformer}, which is a well-known language model for NLP and has the advantages of both \emph{recurrent neural networks} (RNNs) and \emph{convolutional neural networks} (CNNs), to construct the encoder and decoder. The loss function for training the DNN model  is  characterized by two terms: one is the cross entropy which measures the semantic difference between raw and decoded signals, while the other is the mutual information to maximize the system capacity. The SplitNet design was demonstrated to outperform  a traditional communication system in terms of sentence similarity, which is specified in (\ref{eqn:similarity}), and robustness against channel variation. In addition, there are other relevant works on SemCom using the SplitNet approach, e.g., the distributed SemCom system for IoT \cite{Qin2021JSAC} and SemCom system for speech transmission \cite{Qin2021DeepSC-S} (see more details in Section IV-C).

\subsubsection{Comparison between Two Approaches}
The advantages of layer-coupling designs include backward  compatibility, simplicity, and flexibility. Since the approach is based on a modified version of the conventional protocol stack,  SemCom system designed using the approach allows the use of existing coding and communication techniques if they are suitably   modified to allow some control by the Semantic Layer. Furthermore, by modularizing a SemCom system,  individual modules are simpler to design compared to the fully integrated SplitNet. Furthermore, given standarized interfaces between modules, their design can be distributed to different parties. Inevitably, the advantages of layer-coupling approach are at the expense of optimized performance and end-to-end efficiency. Hence, in terms of performance, SplitNet is a better choice. 

A SplitNet design of SemCom system is dedicated to a particular application, somewhat losing the universality of the layered approach. Given a specific task   and a radio-propagation  environment, the encoder and decoder parts  of the neural network 
are jointly trained to efficiently compress raw data into transmitted symbols while ensuring their robustness against channel fading and noise. This makes it possible to achieve a higher communication efficiency and better task performance than a  layer-coupling design. Nevertheless, SplitNet faces its own limitation in three aspects. \emph{First}, channel fading and noise result in stochastic perturbation to both forward-propagation and back-propagation of the DNN. This may result in slow model  convergence during training.  As  proposed in~\cite{Qin2021JSAC},  the issue can be alleviated by feedback of \emph{channel state information} (CSI) to mitigate fading at the cost of additional latency and overhead. \emph{Second}, the radio-propagation environment varies over time and sites, especially for high-mobility applications. A pre-trained end-to-end SplitNet model tends to be ineffective in a new  environment and re-training is needed, which is time consuming and may incur excessive communication overhead. \emph{Third,} analog channel symbols generated by a neural network can be harder for circuit implementation than conventional  modulation  constellations due to, for example, a larger dynamic range. Nevertheless, research on the   SplitNet approach is still in its nascent stage and continuous  research advancements are expected to yield effective solutions for overcoming the above limitations. 

\section{Human-to-Machine Semantic Communications}\label{Section:H2MSemCom}
\begin{figure*}[t]
\centering
\includegraphics[width=0.9\textwidth]{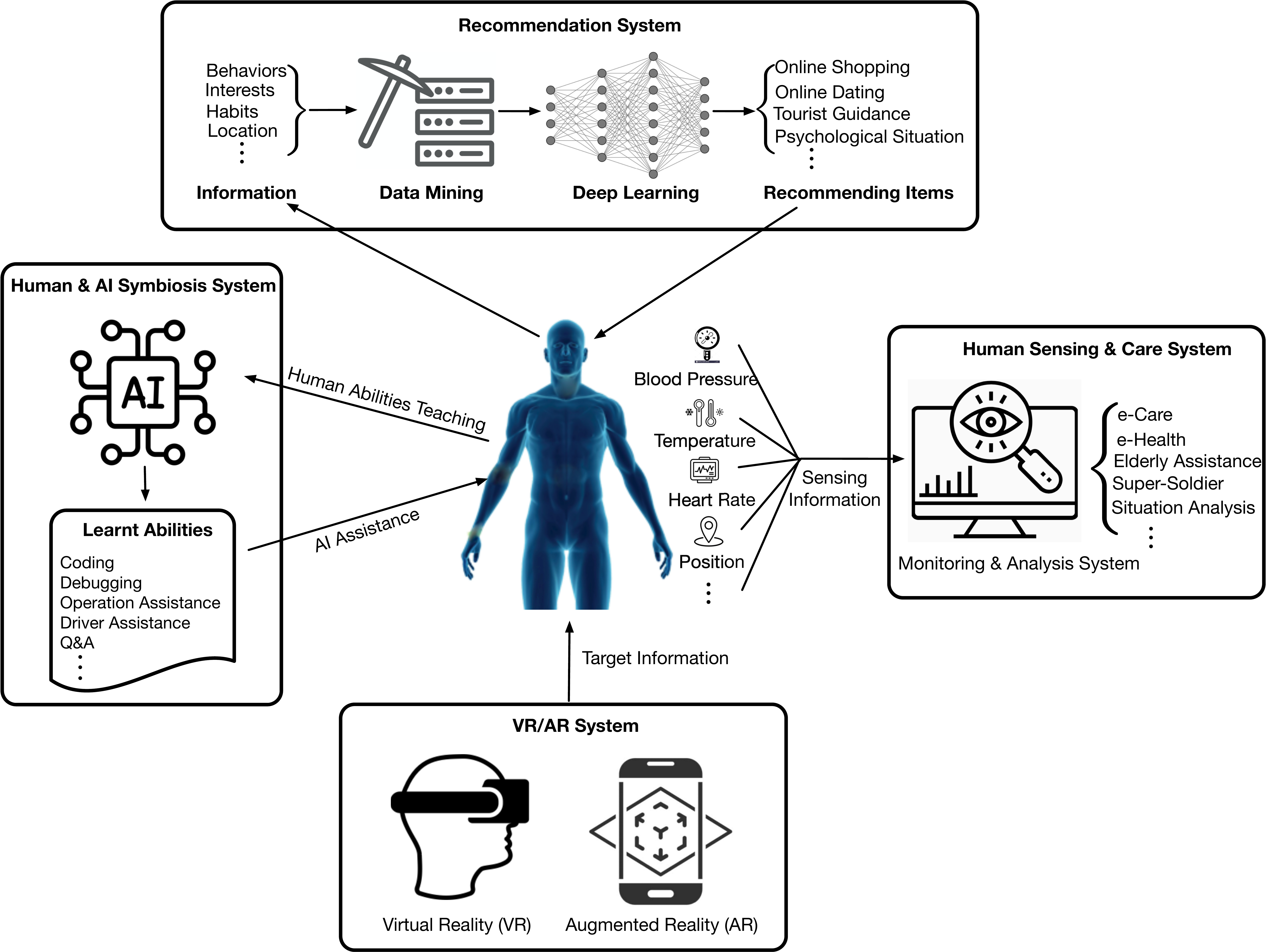}
\caption{Human-to-machine semantic communications.}\label{Fig:HumanMachine}
\end{figure*}

Recall that H2M SemCom features the transmission of messages that can be understood not only by humans but also by machines, such that they can have dialogue or the latter can assist or care for the former. The  potential applications of H2M SemCom are illustrated in  Fig. \ref{Fig:HumanMachine}. In this section, we discuss semantic encoding and other  H2M SemCom techniques in four  representative areas:  human-machine symbiosis, recommendation, human sensing and care, and VR/AR.

\subsection{SemCom for Human-Machine Symbiosis}\label{sec:H2Msymbiosis}
\begin{figure*}[t]
\centering
\includegraphics[width=0.6\textwidth]{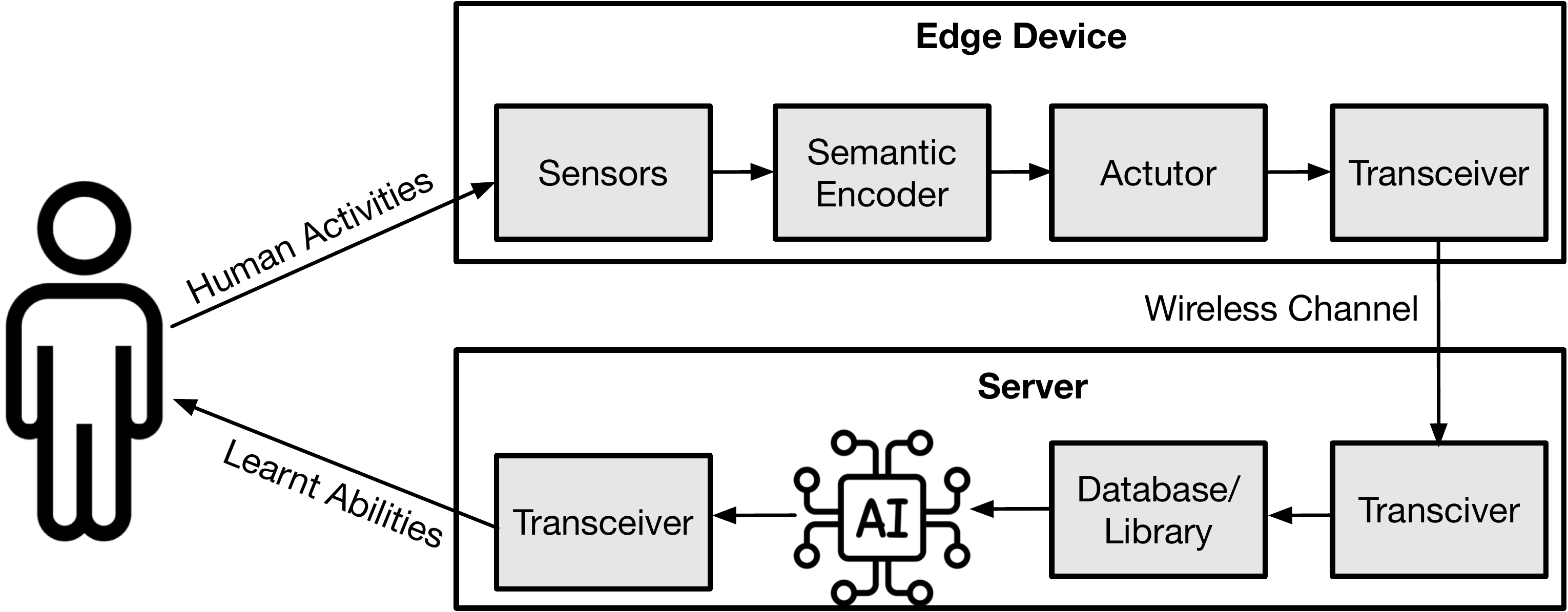}
\caption{Semantic data processing for human \& machine symbiosis systems.}
\label{Fig:Symbiosis}
\end{figure*}

 Human-machine symbiosis (also known as man-computer symbiosis)  refers to  scenarios in which humans and machines establish a complementary and cooperative relationship. On one hand, using their complementary strengths, they can cooperate to carry out a task that is originally difficult or even infeasible.  Humans can benefit from machines' assistance to improve their life quality or productivity. On the other hand, humans and machine can teach each other to improve individuals' capabilities e.g., AI-powered education or imitation learning.  In this sub-section, we discuss SemCom in the scenarios  of Human-machine symbiosis. A typical system is   illustrated in Fig. \ref{Fig:Symbiosis} and its main operations are described as follows. First, human activities are  sensed and the sending results are semantically encoded at  edge devices. Then the encoded data are  transmitted to an edge server for decoding and subsequent use to train an AI model. Finally, the trained AI model acquires some domain specific,  human-like  abilities, which are further used to assist humans. The  distinction of the human-machine symbiosis lies in the semantic encoding techniques that maps human sensing data or knowledge into into low-dimensional vectors while capturing their semantic meanings or latent features \cite{semantic2005}. For example, the encoded  semantic data  refers to the embedded knowledge in the case of  text or  boundaries between objects and the background in the case of image \cite{SegNet,ERNIE1,ERNIE2,SEED}. In the remainder of the sub-section, we introduce two representative  semantic encoding techniques widely used in human-machine symbiosis,  linear LSA models \cite{LSA1,LSA2,ZeroShotLinear} and BERT \cite{BERT,Qin2021DeepSC-S,Qin2021DeepSC}, and discuss their deployment in SemCom systems. Additional  techniques are also  briefly described, followed by an overview of state-of-the-art applications.

\subsubsection{Semantic Encoding by LSA}\label{subsec:LSA}
As a technique in natural language processing, LSA  is used to model and extract semantic information from text documents. LSA has diverse applications, ranging from search engine to translation to the study of human memory. A basic  LSA technique follows  the following procedure.
Consider a set of documents. To begin with, each document  is expressed  as a column vector where each element is binary indicating  whether it includes a specific word or term associated with the element's location. Then putting the column vectors together makes the document set a so called \emph{document-term matrix} denoted as ${\bf X}$. From the matrix, semantic information can be extracted via  the following steps. First,  the principle column subspace of the document-term matrix, called \emph{semantic space}, is computed by SVD: ${\bf X} = {\bf U}{\bm \Sigma}{\bf V}^T$, where ${\bf U}$ is the column space of ${\bf X} $, ${\bm \Sigma}$ is the singular value matrix, and ${\bf V}^T$ is the row space. 
Given desired dimensions $k$,  the semantic space is defined as ${\bf U}_k$, which is the $k$-dimensional principle column subspace of ${\bf X}$. The corresponding $k$-dimensional principle singular-value matrix is denoted as  ${\bm \Sigma}_k$. Second, all document vectors are projected onto  the low-dimensional semantic space. Specifically, let  ${\bf d}_j$ represents  the $j$-th column of ${\bf X}$ and is thus  the $j$-th document. Then the extracted  semantic vector, denoted as $\hat{\bf d}_j$, is obtained as 
$\hat{\bf d}_j = {\bm \Sigma}_k^{-1}  {\bf U}_k^T {\bf d}_j$.
Last, using the reduced-dimension semantic vectors, the similarity level between any two high-dimensional  documents, say $j$-th and $j^{\prime}$-th documents,  is measured by efficiently  computing the following function: 
\begin{equation}
    S(j,j^{\prime}) = \dfrac{\hat{\bf d}_j\cdot \hat{\bf d}_{j^{\prime}}}{\left\| \hat{\bf d}_j \hat{\bf d}_{j^{\prime}}\right\|}, \quad \forall j\neq j^{\prime}.
\end{equation}

In the context  of SeCom between a human and a machine, the function of LSA is to extract semantic information from human speech or messages and in that way aid the machine's interpretation. Its deployment in a SemCom system essentially involves the  design of a LSA-based semantic encoder and substituting the result into either the layer-coupling architecture (see Fig. \ref{Fig: layered architecture}) or the SplitNet architecture (see Fig. \ref{Fig: JSCC}). As an example, the design proposed in \cite{jiang2021} is based on SplitNet and features integrated  semantic/channel coding/decoding implemented by training a split DNN model with the transmitter half  performing  LSA. On the other hand, for a design based on the layer-coupling architecture, LSA resides in the semantic layer to map each human message into the semantic space. The distilled semantic information  in the dimension associated with a  larger singular value is more important. This suggests the use of the singular values as DII indicators. Then the LSA-encoded information, together with the DII indicators, are passed to the lower  layers for transmission. On the other hand, the CRI feedback in the upward direction enables the semantic layer to  adapt the dimensionality of the semantic space to the channel state. Consequently, when the channel supports a high rate, the   semantic space can be expanded to yield  a better representation of the human message and thus a more accurate understanding by the machine at the receiving end.

\subsubsection{Semantic Encoding by BERT}\label{subsec:BERT}
BERT is a well known language processing approach based on a popular model called \emph{transformer}, which is the first transduction model relying entirely on self-attention to compute representations of its input instead of  using sequence-aligned RNNs or convolution to generate its output~\cite{vaswani2017attention}. A transformer comprises an encoding component and a decoding component.  Each  includes several sequentially connected encoders or decoders. An  encoder cascades  one self-attention layer with  a feed forward neural network. 
The former performs feature extraction to find  the relation of words in  the input sentence; the latter is trained with a suitable  objective, such as language translation. A decoder has a similar structure as the encoder except for having an extra encoder-decoder attention layer inserted between the self-attention layer and the feed forward neural network. The additional layer helps the decoder focus on a specific position in the input sequence to handle issues, such as the case of one word having multiple meanings.

 Building on the  transformer architecture, the key feature of BERT is a new training strategy, termed \emph{masked language model} (MLM), that randomly masks some words in a sentence to generate training samples. The training objective is to learn the masked words in the sentence.  This training strategy and objective make it possible to generate an enormous number of unlabelled text data samples for  training. Another key feature of BERT is that a text sentence is input into the transformer as a whole rather  word-by-word following the natural left-to-right uni-direction. The features endow on the trained transformer  the ability of predicting the missing words based on their context, giving the technique the name of bidirectional representations. With this ability, BERT outperforms the  uni-directional approaches to become the state-of-the-art strategy for natural language processing. The combination with  other techniques (e.g., classification) broadens the applications of  BERT,  e.g., Q\&A, and  information retrieval.  
 
The procedure of deploying BERT in a SemCom system to support human-machine symbiosis is similar to that for LSA. In other words, BERT can be simply used as the  semantic encoder. For the SplitNet design, BERT is used as a part of the split  DNN. For the layer-coupling approach, BERT is used for extracting the important information from the sensed human activities with DII indicators showing their importance levels. The   exchange of data and control signals between the semantic and lower layers  are similar to those  for LSA. 

\subsubsection{Additional Semantic Encoding Techniques}
Other techniques that can play an important role in semantic encoding for human-machine symbiosis are the  CNN based approaches of  object recognition \cite{SemanticSegmentation1,SemanticSegmentation2,SemanticSegmentation3}. Relevant techniques can efficiently compress human sensing data (e.g., facial expressions and behaviours) for efficient transmission to machines for subsequent recognition. A typical object recognition technique detects the boundary between the targeted objects and the background based on contextual features. A representative design of CNN-based \emph{auto-encoder} (AE) for segmentation is proposed in \cite{SegNet}, termed SegNet, which comprises an encoding component, a decoding component, and a classifier. The encoding component contains several encoders, each of which is paired with a decoder in the decoding component. Each encoder is a CNN, modified from  the well known  VGG-16 network  \cite{ImageRecognition}. Each decoder  up-samples its input using the transferred pool indices from its corresponding encoder to produce sparse feature maps. Finally,  the output feature maps of the last  decoder are fed to a softmax classifier for pixel-wise classification. This generates the segmentation results.

\subsubsection{Choice of the Connectivity Type}
In 5G systems, there exist three generic connectivity types: enhanced Mobile Broadband (eMBB), Ultra-Reliable Low-Latency Communication (URLLC), and massive Machine Type Communication (mMTC). They are defined to support a wide range of services with heterogeneous \emph{quality-of-service} (QoS) requirements. The choice of connectivity type for human-machine symbiosis is application dependent. Many relevant  applications do not require high transmission rates, ultra reliability,  or  massive connections. Examples include  AI-assisted learning, coding, and debugging. For such applications, normal  radio access suffices. On the other hand, there exist a class of symbiosis applications that involve tactile interaction, thereby requiring low latency and reliable transmission. For instance,  AI-assisted driving and remote surgery require the response  latency between robots and humans (i.e., drivers or surgeons)  to be less than $10$ ms and $1$ ms, respectively \cite{jiang2020ai}. For this class of applications, the provisioning of URLLC connectivity is crucial. 

\subsubsection{State-of-the-Art Applications}
Applications related to human-machine symbiosis can be separated into three main classes.  The first class  of applications is  \emph{AI-assisted systems}. AI technologies provide ways for machines to  acquire human-like skills and abilities by learning from the experiences of human experts (e.g., doctors and drivers), which, in turn, make the machines useful  assistants for humans. Using LSA, AI-powered  chatbots have started to replace humans in  FAQs/customer  services in places such as universities \cite{xu2017new} and over  social media \cite{ranoliya2017chatbot}. Machines can also  play an important role in AI assisted healthcare via  utilizing machine learning algorithms to extract key information from patients' records to help doctors with diagnosis  and prediction of  the risks of diseases  \cite{lysaght2019ai}. Machine assistance have also been applied to  other professional areas, such as video game debugging \cite{machado2018ai}, automatic programming  \cite{svyatkovskiy2019pythia}, driving assistance \cite{ziebinski2017review}, and second language learning and teaching  \cite{kannan2018new}. 

The second class of applications is \emph{interactive machine learning} that includes humans in the loop to leverage the generalized  problem-solving abilities of human minds  \cite{holzinger2013human,holzinger2016interactive,berg2019ilastik, schneider2020human, de2013exploring}.  This is particularly useful in  cases  lacking training samples for  rare events that are needed for automatic machine learning to work. Moreover, the joint force of machines and humans can  combine their complementary strengths to tackle grand challenges such as protein folding and $k$-anonymization of health data \cite{holzinger2016interactive}. In such collaboration, human experts  uses their  experiences to guide  machines  to reduce  the search space.

The third class of applications is  \emph{worker-AI collaboration}  where both human and machine workers  cooperate as peers to finish real-time tasks, e.g., moderating content, data deduplication \cite{sriraman2017worker,AMT}). In particular, relying on  tactile communication,  robots  can imitate the actions of remote surgeons  in minimally invasive surgery \cite{okamura2009haptic, sornkarn2016can}. Such cooperative surgeries can benefit from the machine involvement to improve the accuracy and dexterity of a surgeon and minimize  traumas induced on  patients. One important design issue  for   worker-AI collaboration is to prevent machines from telling lies or making mistakes \cite{chakraborti2019can}.

\subsection{SemCom for Recommendation}
 A  recommendation system predicts user preferences in terms of ratings of a set of items such as songs, movies, and products. Recommendation has become a popular tool for making machines intelligent assistants and improve user experience. Examples include playlist generation  for multimedia streaming  services, product recommendation  for online shopping, marketing  on  social media,  Internet search, and  online dating \cite{rosa2018knowledge,xia2015reciprocal,zhao2018analyzing,tang2018novel}. A SemCom system aims at supporting recommendation in a wireless network, as shown in Fig.~\ref{Fig:HumanMachine}. In the system, an edge device semantically encodes and transmits the user's personal data  to a edge server for generating recommendations  for the user. The purpose of semantic encoding is to infer user ratings from the user data. Given a rating database of a large number of users, the server generates recommendations for a target user using a filtering technique. Among the most popular one  is collaborative filtering discussed in the sequel. We will also introduce other techniques including  content-based, collaborative, and hybrid filtering. The choice of connectivity type for SemCom systems to support recommendation will be also discussed, followed by an overview of the state-of-the-art applications of recommendation systems.

\begin{figure*}[t]
\centering
\includegraphics[width=0.6\textwidth]{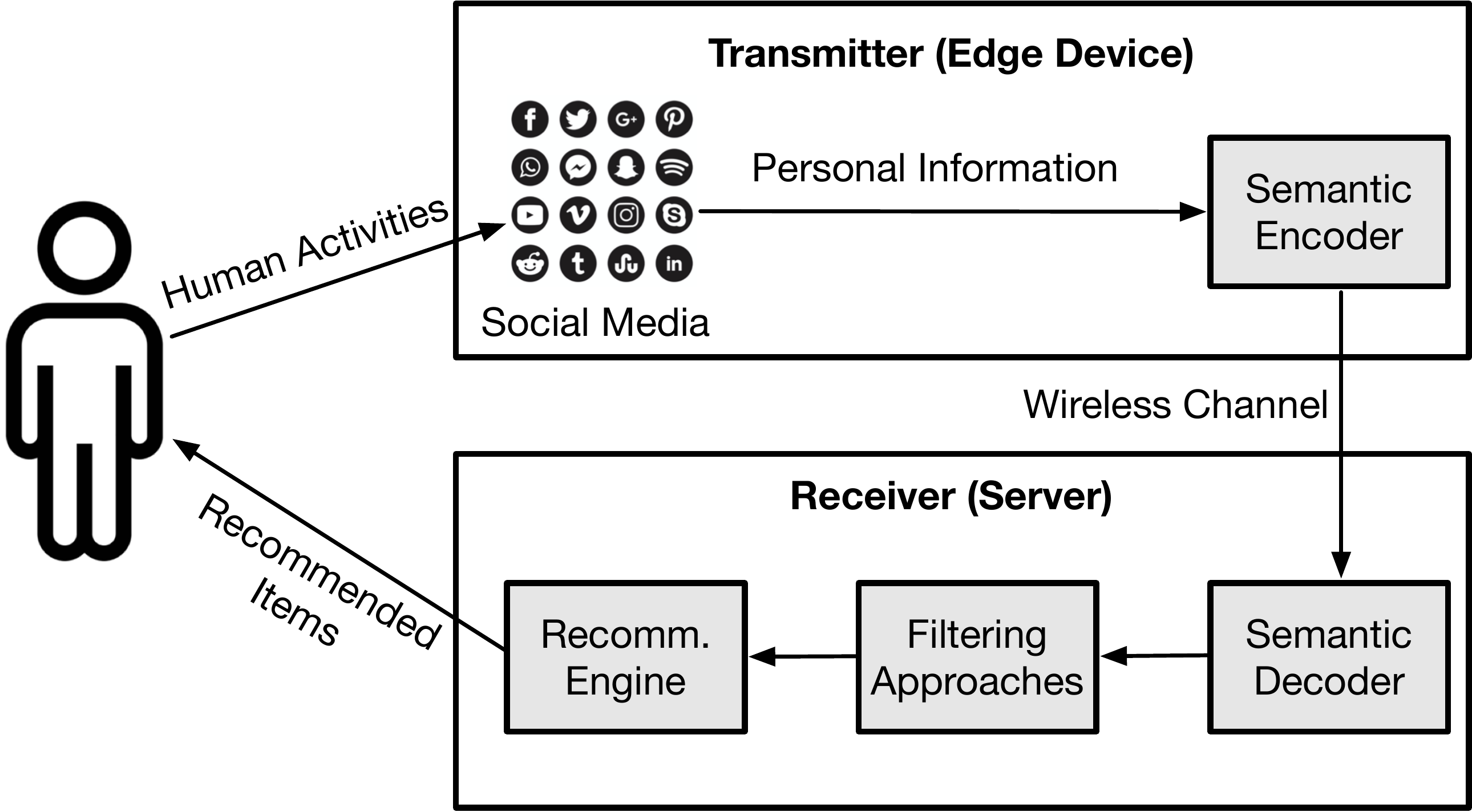}
\caption{Semantic data processing for recommendation systems.}
\label{Fig:Recommendation}
\end{figure*}

\subsubsection{Collaborative Filtering} 
Collaborative filtering finds users with similar preferences  using their historical ratings \cite{yang2016survey}. Then the purpose of semantic encoding  is to distill the rating information from the historical data recording user's daily activities such as shopping, entertainment, reading, and multimedia streaming. This removes redundant information to compress the data for efficient transmission. The design of semantic encoder can be based on LSA described in Section III-A by modifying the document-item matrix to count the user's access/purchase frequencies of different items. In the case where such explicit information is unavailable, an AI model  can be trained  to infer the users' preferences from sensing data recording his/her behaviour and emotions in either the physical world or on social-media platforms~\cite{wang2015collaborative}.

Next, after receiving rating data from multiple users, the server compiles them into a user-item matrix. Each column storing the ratings by  one specific  user is called an item vector. Let ${\bf r}_j$ denotes the $j$-th item vector and ${\bf r}_{j,i}$ its $i$-th element representing the rating of $i$-th item in a set of interest. 
Moreover,  $\bar{\bf r}_j$ denotes the average rating of all items of user  $j$. Using the Pearson correlation as  a metric, the similarity in preference between user $j$ and $i$ can be computed as 
 \begin{equation}
     S(j,j^{\prime}) = \dfrac{ \sum\limits_{i\in I_{j,j^{\prime}}} ( {\bf r}_{j,i} - \bar{\bf r}_j) ( {\bf r}_{j^{\prime},i} - \bar{\bf r}_{j^{\prime}}) }{ \sqrt{  \sum\limits_{i\in I_{j,j^{\prime}}}  ({\bf r}_{j,i} - \bar{\bf r}_j)^2 } \sqrt{  \sum\limits_{i\in I_{j,j^{\prime}}} ( {\bf r}_{j^{\prime},i} - \bar{\bf r}_{j^{\prime}})^2 } },
 \end{equation}
 where $I_{j,j^{\prime}}$ represents the set of items rated by both user  $j$ and $j^{\prime}$, The pairwise similarity  measures allow the server to recommend items preferred  by some  users to others sharing similar interests. 
 
 However, as social-media applications are fast growing in number and type, the user-item matrices become increasingly  sparse. The insufficient rating data causes difficulty in clustering of  similar users. Researchers have developed solutions for this problem by applying  techniques from  data mining and machine learning including  SVD \cite{zhou2015svd}, non-negative matrix factorization \cite{huang2016collaborative}, clustering \cite{zahra2015novel}, and probability matrix factorization \cite{chaney2015probabilistic}). 

As in the scenario of human-machine symbiosis, SemCom systems for recommendation can be based on either the layer-coupling or SplitNet architectures. When explicit rating information is available at a device, its uploading is infrequent and may even require only a single upload, as user preferences usually do not change rapidly over time. On the other hand, when such explicit information is unavailable, a large amount of user sensing data may need to be transmitted from the device to the server for preference inference. One way to address this issue is by designing semantic encoders that can locate a low-dimensional item-rating subspace without compromising the recommendation  accuracy. The other way is utilizing high-rate access (eMBB) whenever it is available or by deploying a targeted large-rate technology, such as mmWave.

\subsubsection{Other Filtering Techniques}
Other available  filtering techniques for recommendation include content-based filtering, demographic filtering, and hybrid filtering \cite{ryngksai2014recommender}.  The content-based approach utilizes the users' historical data  for recommendation. Specifically, the recordings of, e.g., habits or interests, are useful for creating a user profile characterized by a set of features. Then, an item aligned with the  features of a profile is likely to interest the associated user and thus can be recommended. Next,  the demographic filtering approach classify users according their demographic information such as nationality, age, and gender. Items preferred by one user are recommended to the users in the same demographic class. Last, the hybrid filtering approach combines several  aforementioned approaches and has been showcased to boost the recommendation  accuracy.

\subsubsection{State-of-the-Art Applications}
Recommendation systems are deployed in many areas. The most popular venue is social networks where recommendation is applied to emotional health monitoring by detecting abnormality \cite{rosa2018knowledge},  partner recommendation in online dating \cite{xia2015reciprocal}, and emoji usage suggestions  \cite{zhao2018analyzing}. 
Other applications include travel recommendation systems for mobile tourist \cite{gavalas2011web}, remote healthcare (e.g., cloud-assisted drug prescription   \cite{zhang2015cadre} and cloud-based mobile health information in \cite{wang2016design}), TV channel  recommendation \cite{chin2010cprs}, video recommendation  \cite{davidson2010youtube}, and  music recommendation \cite{ayata2018emotion}. Traditionally, to offload the high computation load, recommendation systems  are hosted in  the cloud server with unlimited computation resources \cite{zhang2015cadre,wang2016design,mo2014cloud}. Nevertheless, the traditional approach can lead to excessive  communication latency and overhead as the personal data to upload is known to grow  at an exponential rate. Recent years see the increasing popularity  of the split-computing approach that spreads a recommendation system across the cloud and the network edge leveraging the edge computing platform \cite{wang2020edge}. In addition, researchers have proposed \emph{unmanned aerial vehicle} (UAV) assisted  recommendation systems  for location based social networks  \cite{tang2018novel} as well as distributed recommendation systems featuring  data privacy  \cite{corchado1995distributed,armknecht2011efficient}.


\subsection{SemCom for Human Sensing and Care}\label{sec:H2Msensing}

Human sensing-and-care refers to real-time tracking and monitoring of humans' health conditions and movements by machines, such as the machines can offer a proper care to the humans. The human monitoring relies on sensors (e.g., temperature and positioning) on or around humans. The sensing data are then transmitted to a server for analysis and decision making. To facilitate the discussion, let us consider the concrete example of biomedical sensors, which are wearable or implantable and perform transduction of biomedical signals (e.g., \emph{electrocardiogram} (ECG) signals) into electric signals. In this context, the purpose of designing a SemCom system is extract useful features from biomedical sensing data  and transmit them to a server for  diagnostics or  medical image analysis. An example of a feature is the ``main-spike"  interval of  ECG signals, termed QRS interval. In the sequel, we discuss the techniques for biomedical semantic encoding and the associated SemCom system design.

\subsubsection{Biomedical Semantic Encoding}

Typical biomedical signals include ECG signals for detecting heart activity  and \emph{electromyography} (EMG) signals for detecting e.g.~skeletal activity. Such signals are characterized by a certain level of periodicity  and predictability, making it possible to estimate the signal statistics within a short time frame. In the current context,  semantic coding is particularized to techniques for estimating the statistics that contains useful information the biological activities of a human. Relevant techniques are based on either the time-domain or  frequency-domain approaches.  As an example, consider the R-peak detection of a ECG signal, where R refers a point corresponding to a  peak of the ECG wave. The detection of R-peak helps the heart-rate characterization \cite{oweis2014qrs}. More elaborate analysis of a ECG wave decomposes  a main spike into three successive   upward/downward  deflections, termed Q wave, R wave, and S wave.  A  time-domain method for their  detection  mainly use the shape characteristics, such as finding the largest first-order and second-order derivatives. On the other hand, a frequency domain method first transforms the signal into the frequency domain using,  e.g., wavelet transformation, and then apply filtering with a suitable  passband to  extracting the desired information.

In the SemCom system for human sensing-and-care, the biomedical signals that are semantically encoded and transmitted by an small-size edge device. Upon detecting abnormality, the transmissions from this device should be real-time and very reliable, to call for an urgent medical care~\cite{pan1985real}. In view of these requirements, it is preferable to design the SemCom system  on the layer-coupling architecture rather than a DNN model. This is because no complex data should be processed and the complex DNN model may be an overkill, while its long computation time  unacceptable. The semantic encoding and transmission can be controlled using DTI generated as follows. The duration, amplitude, morphology, frequencies  of Q/R/S waves are  all useful for heart related  diagnostics ranging from detecting  conduction abnormalities to diagnosing ventricular hypertrophy. But they are of different importance levels that are also disease-dependent. In other words, the  features of biomedical signals can be assigned different importance levels, resulting in the DII. At the lower layers, the adopted radio-access technologies  depend on applications. For indoor applications, sensors are usually linked  to a local hub (e.g., a smartphone)  using short-range and low-latency  technologies such as Zigbee, Bluetooth, and WiFi. For the hubs to access the cloud or for outdoor applications, cellular communication is the preferred  choice. In regions with no or poor cellular coverage, satellite communications can be used instead while GPS helps human positioning and tracking. A large-scale network that  connects a massive number of sensors  can rely on the mMTC service supported within the 5G architecture.

\subsubsection{State-of-the-Art Applications}
The common application of human sensing-and-care is elderly monitoring \cite{suryadevara2012wireless,lin2006wireless}. In  \cite{suryadevara2012wireless}, a wireless sensor network is deployed to monitor the well-being conditions of  the elderly. Specifically,  multiple types of sensors are used to monitor their activities such as cooking, dining, and sleeping. A similar system targeting the elderly  with dementia is reported   in \cite{lin2006wireless}. Another type of application is a super soldier system~\cite{javaid2013measuring}, which monitors  and analyzes the health status and fatigue levels of soldiers by sensing their temperatures, gestures,  blood glucose levels, and ECG. The third type of application is the set of general human activity recognition systems \cite{windau2013situation,golestani2020human}. In \cite{windau2013situation}, head-mounted smartphones are designed to have situation awareness, e.g., awareness of user behaviors and environmental conditions. To this end, the data collected from smartphone sensors (e.g.,  accelerometers, gyroscopes, and cameras) are transmitted to a server for feature extraction and situation inference. In another design presented in \cite{golestani2020human},  a wearable magnetic induction device is used for sensing and wirelessly transmitting the magnetic induction signals to a server for activity detection using  a RNN based algorithm. Other applications of human sensing-and-care include remote healthcare systems  \cite{liszka2004keeping,WP1,WP2} and smart-home monitoring systems \cite{kelly2013towards}).

\subsection{SemCom for VR/AR}\label{sec:H2MVRAR}
VR and AR are  two H2M technologies. VR essentially involves the use of mobile devices  (e.g.,  smartphones, glasses, or  headsets) to create new human experiences by replacing the physical  world with a virtual one. On the other hand, AR devices  alter humans experiences by augmenting real objects with computer-generated perceptual information across different senses (e.g., vision, hearing, haptics, hearing, pressure, and smell). VR/AR provide a way of seamlessly merging the physical and virtual words. The resulting immersive human experience can give a rise to a plethora of future services, such as  entertainment, virtual meetings or remote education. Offloading computation and caching to edge servers makes it possible to implement  latency sensitive VR/AR applications on  resource-limited devices. VR/AR data processing and SemCom between  devices and servers are discussed in the following sub-section followed by a summary of state-of-the-art SemCom for AR/VR.

\subsubsection{VR/AR Semantic Encoding and Transmission}
\begin{figure*}[t]
\centering
\includegraphics[width=0.8\textwidth]{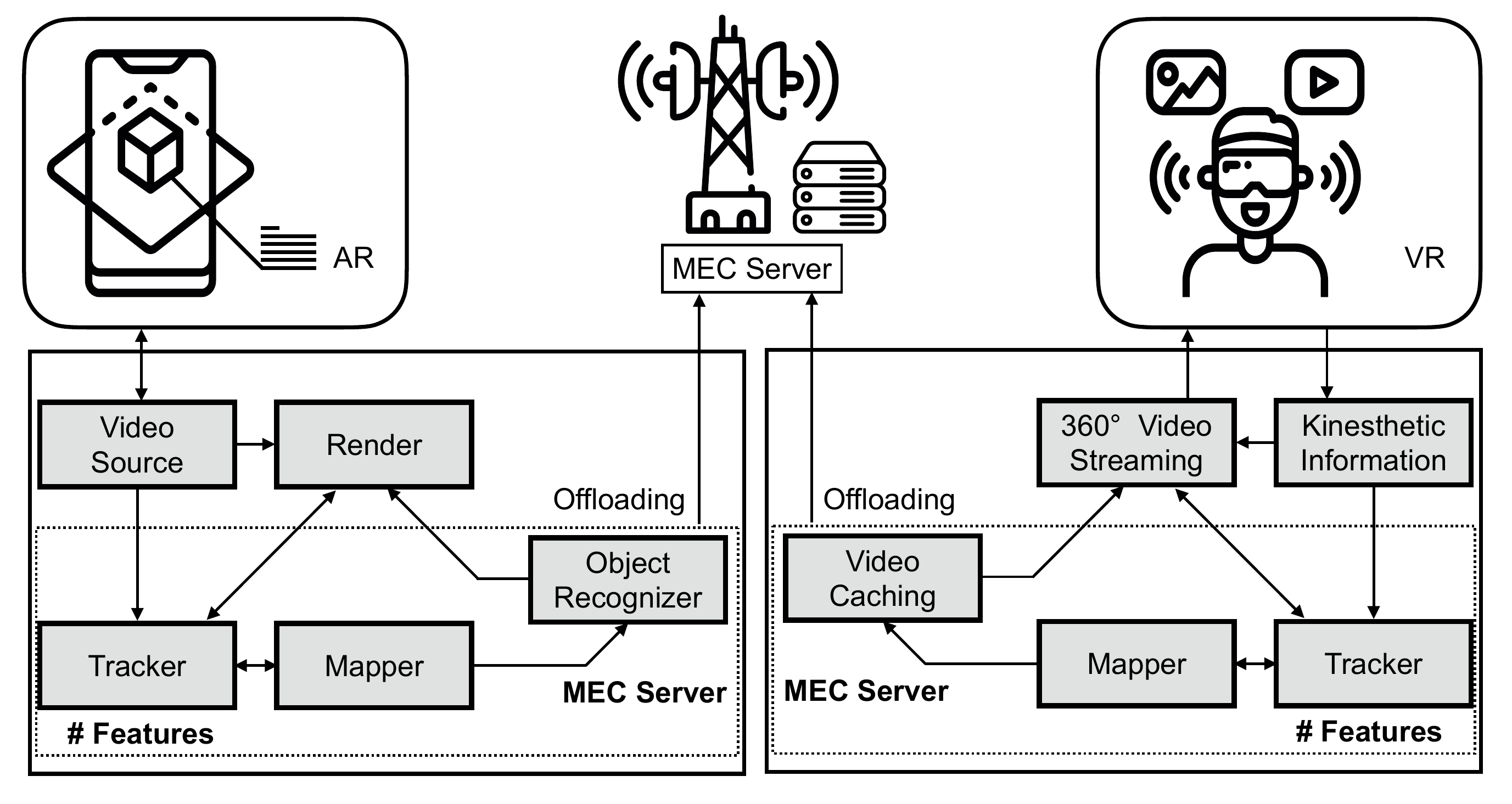}
\caption{VR/AR systems enabled by MEC.}
\label{Fig:VRAR}
\end{figure*}

The procedures of semantic encoding and transmission in AR and VR systems are illustrated  in  Fig.~\ref{Fig:VRAR}. Consider AR semantic encoding whose purpose is to recognize and track physical objects of interest to the user and then project icons,  characters and information  onto them. Its implementation requires cooperation between a device and an \emph{mobile edge computing} (MEC) server. 
First, raw video data recorded locally using on-device cameras are uploaded to the  server for processing. In the MEC server, three algorithms, namely \emph{mapper, tracker, and object recognizer},  are executed~\cite{VRARRen2019}. The function of the tracker is to detect the object's position based on input raw data and to proactively adjust a rendering focal area. Based on the tracking results, the mapper is to distill features (e.g., virtual coordinates) of objects embedded in the raw data using image processing techniques. In parallel, the object recognizer leverages both the object features and video streaming to produce desired  rendering data (e.g., cartoon icons and explanatory text) according to the application requirements. Such data are downloaded onto the device where they are superimposed onto the actual scenes by a local renderer and the edited VR videos are displayed to the human user. Next, the function of VR semantic coding is to select only part of the video depicting  the virtual world to download and display to the user such that the heavy burden of downlink transmission is alleviated~\cite{elbamby2018toward}. To this end, the user's kinesthetic information (e.g. location, angle of view, and head movements) is collected over multiple on-device sensors and efficiently transmitted to the server. The information is processed by a tracker and a mapper operating at the server to detect the \emph{field-of-view} (FoV) and select the corresponding video output by extraction from  cached 360$^{\circ}$ video streaming such that it best fits the user's movements. Then video output is downloaded onto a pair of VR glasses or a VR  headset for constructing the virtual world. Such semantic encoding dramatically reduces the required downlink data rate as opposed to the full 360$^{\circ}$ video streaming. It also makes it possible to meet the stringent latency requirement for immersive  user experience. The communication efficiency can further improved by deploying advanced semantic encoding techniques such as video segmentation and compression by head-movement prediction and eye-gaze tracking. 

The connectivity requirements for VR/AR SemCom are discussed as follows. In general, VR/AR systems need to collect and process real-time multimedia  data from the physical world and generate/transmit  high-resolution visual and auditory data. Therefore, the required connectivity is characterized by high rate and low latency~\cite{elbamby2018toward,chen2018virtual,yu2019skin}. As an example,  a  human FoV covers a horizontal and a vertical ranges of  150$^{\circ}$ and 120$^{\circ}$, respectively. The simulation of a realistic FoV generally requires $120$ frames per second with each frame consisting of $64$ million pixels (60 pixels/degree). Given standard video quantization ($36$ bit/pixel) and H.$265$ encoding (with $1:600$ compression rate), the required transmission rate  is at least $1$ Gb/s~\cite{elbamby2018toward,Samsung20206GVision}. On the other hand, real-time  interaction needed for immersive human experiences requires motion-to-photon latency to be lower than $15$ ms~\cite{HuaweiXRlatency2018}.  Such requirements places  VR/AR connectivity  at the intersection between eMBB and URLLC. A solution that addresses these issues is the MEC platform in 5G that offloads computation intensive  tasks (e.g. tracking, mapping, and recognition) and caching storage-demanding multimedia content at edge servers in the proximity of users as shown in Fig.~\ref{Fig:VRAR}.  This reduces the burden of devices to be  merely responsible for data collection and displaying videos. 

Building the MEC platform, a  VR/AR system can be designed based on either the  layer-coupling or the SplitNet architecture. Consider the former. Different types of human kinesthetic information are of heterogeneous importance for a specific application and can be thus assigned different DIIs to facilitate importance aware adaptive transmission.  On the other hand,  for  collaborative VR/AR involving multiple  devices and servers, the SemCom system design can benefit from exploiting the  PAI of AI models (e.g. classification) and other data processing algorithms (e.g. compression and filtering), the DTI and DII of raw data (e.g. voice and images) to optimize  the operations of data aggregation and rendering data feedback for boosting the  communication efficiency. On the other hand, the scene-data collection, local rendering, and global data processing at servers can be integrated in an end-to-end design using the SplitNet approach. Then the  trained neural-network is split for partial implementation at a device and server according to the application requirements and device's resource constraints.

\subsubsection{State-of-the-Art SemCom for VR/AR}
There exists a wide range of  VR/AR applications with a vast literature (see e.g., the surveys in~\cite{VRARsurvey1,VRARsurvey2} and references therein). However, the area of SemCom for VR/RA is relatively new and still largely uncharted. Some recent advancements are highlighted as follows. The challenges and enablers for URLLC communications to implement VR/AR are discussed  in~\cite{elbamby2018toward}. Furthermore, a case study of deploying VR in wireless  networks is also presented, which integrates millimeter-wave communication, edge computing, and proactive caching. Another design of  wireless VR network is proposed  in~\cite{chen2018virtual}. It is proposed that  small base stations are used to first collect and track information on  a  VR user and then send to the user device the generated 3-D images. The  resource management issue targeting such a system is also investigated that accounts for VR metrics such as tracking accuracy, processing delay, and transmission delay. In addition, a new type of VR/AR system enhanced by  skin-integrated haptic sensing is proposed in~\cite{yu2019skin}. Such a special wireless sensor can be  softly laminated  onto the curved skin surfaces  to wirelessly transmit  haptic information conveying  the  spatio-temporal  patterns of localized  mechanical vibration.

\section{Machine-to-machine Semantic Communication}\label{Section:M2MSemCom}
\begin{figure*}[t]
\centering
\includegraphics[width=0.9\textwidth]{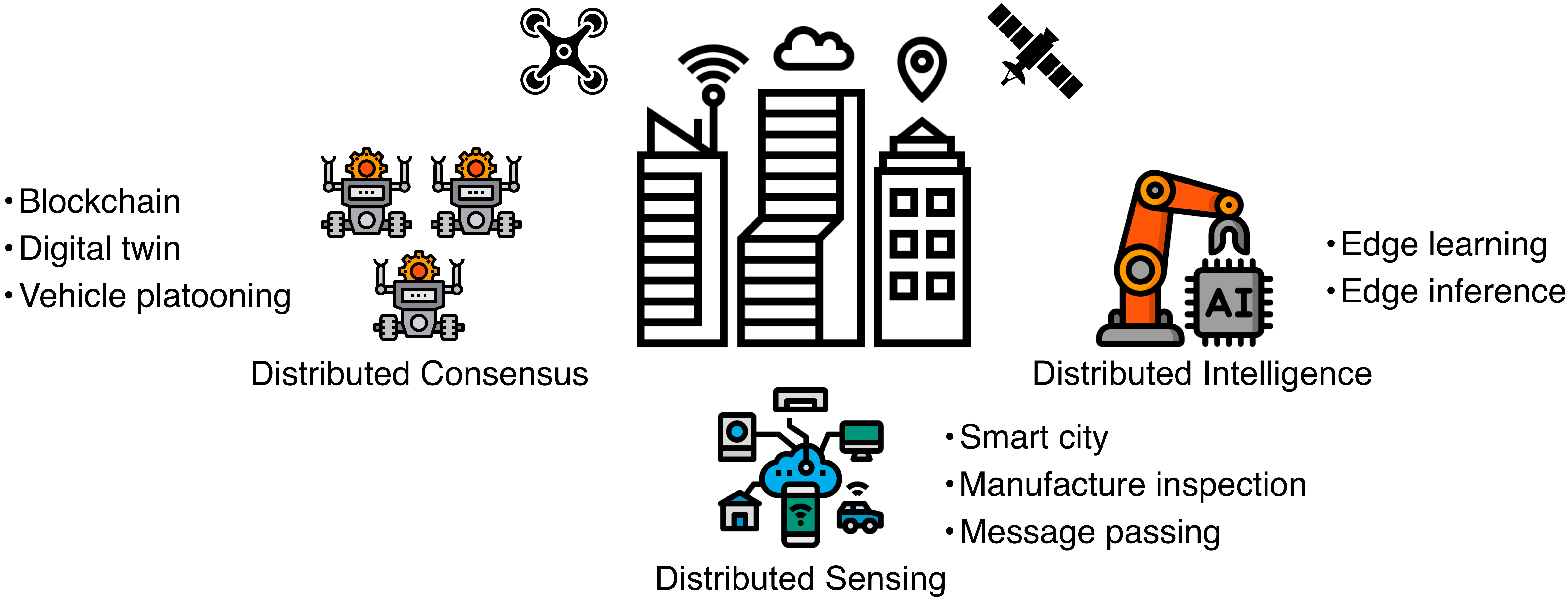}
\caption{IoT application areas relevant to of M2M semantic communication.}
\label{fig: M2M_Applications}
\end{figure*}

Recall the objective of  M2M SemCom is to efficiently connect multiple machines and enable them to effectively execute  a specific  task in a wireless network. It usually targets  IoT applications as illustrated in Fig.~\ref{fig: M2M_Applications}.  The typical tasks in M2M SemCom span the areas of sensing, data analytics, learning, reasoning, decision making, and actuation \cite{Popovski2020JIIS}. In this section, we discuss effectiveness encoding and transmission techniques in four representative types of  application, namely distributed learning, split inference, distributed consensus, and machine-vision cameras. 

\subsection{Distributed Learning}\label{sec:M2MDL}
The main theme of distributed machine learning is to train an AI model using distributed data at many mobile devices as well as their computation resources. 

\begin{figure*}[t]
    \centering
    \includegraphics[scale=0.2]{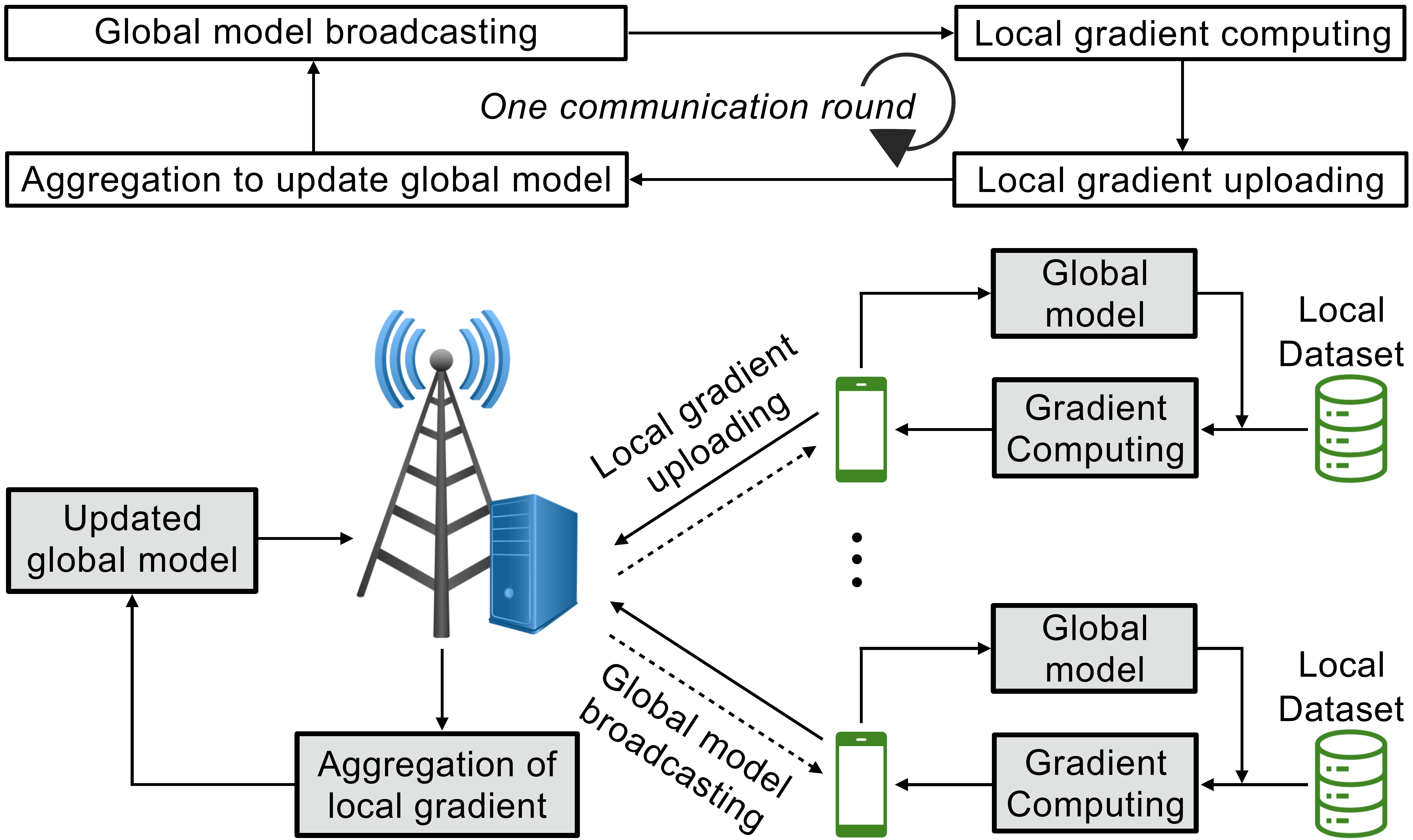}
    \caption{Federated learning in a wireless system.}
    \label{fig: M2M_FL_diagram}
\end{figure*}

FL mentioned in Section~\ref{subsec: effectiveness_encoding} stands out as arguably the most popular distributed-learning framework~\cite{Chen2020Tutorial,Niyato2020Tutorial}. Its popularity is mainly due to its feature of protecting the ownership of mobile data by avoiding their direct uploading to a server. Instead, based on the classic \emph{stochastic gradient descent} (SGD) algorithm, FL requires each device to compute a local model updated using local data or a stochastic gradient representing the update, as illustrated in Fig.~\ref{fig: M2M_FL_diagram}. Then the local model updates are transmitted to the server for aggregation before updating the global model. The aggregation operation suppresses the noise in local updates arising from the limited size of local data. As a result,  the noise  diminishes as the number of devices grows. Subsequently, the server broadcasts the updated global model to all devices to repeat the above process and the iteration continues until the global model converges. While SplitNet targets inference using a trained model,  the layer-coupling approach is a more suitable approach for designing a  FL system. In a FL system, the uploading of high-dimensional model updates by many devices poses a communication bottleneck. For instance,  the popular  ResNet-50 model comprises $25.6$ million parameters or equivalently $1638.4$ million bits in  the ``float64" format. Relevant SemCom techniques for  tackling  this bottleneck  including effectiveness encoding, modulation, multi-access, and \emph{radio-resource management} (RRM) are discussed separately in the following sub-sections.

\subsubsection{Effectiveness Encoding} 

Consider a FL system with one server and $M$ devices, called workers, and an arbitrary  \emph{communication round} (i.e., iteration) in the FL algorithm, say the $t$-th round, that comprises several sequential phases: model broadcast, local effectiveness  encoding, model-update uploading,  and global-model updating. The focus of this subsection is the effectiveness encoding at a device. Its goal   is to convert local training  data into a \emph{local model} by updating the broadcast global model, or a \emph{local (stochastic) gradient} representing the update. At the beginning of the $t$-th round, the server broadcasts the global-model parameters $\mathbf{W}^{t}$ to all workers. Its local gradient as computed at  worker-$m$ and the worker's local dataset are denoted as  $\mathbf{G}_{m}^{t}$ and  $\mathcal{D}_k=\left\{\left(\mathbf{x}_{m,i},y_{y,i}\right)\right\}_{i=1}^{N_m}$, respectively,  where $N_m$ is the local-dataset size, $\mathbf{x}_{m,i}$ the $i$-th sample, and  $y_{m,i}$ its  associated label. The effectiveness encoding involves multi-step (say $B$-step) local gradient descent. To this end, let $\mathcal{D}_m$ be partitioned into $B$ mini-batches with the $b$-th mini-batch denoted as $\mathcal{D}_{m,b}$. 
The  local  gradient in step $b$ is computed  as
\begin{equation} 
\label{eqn: back-propagation}
\mathbf{G}^{t,b}_m = \frac{1}{\vert\mathcal{D}_{m,b}\vert} \sum_{\left(\mathbf{x}_{m,i},y_{m,i}\right)\in \mathcal{D}_{m,b}}\nabla\mathcal{L}\left(\mathbf{x}_{m,i},y_{m,i};\mathbf{W}^{t,b-1}_k\right),
\end{equation}
where $\mathcal{L}(\cdot)$ denotes the \emph{loss function} pre-defined for the learning task. The above computation can be  implemented using  the well-known \emph{back-propagation} algorithm. Essentially, implementing the differential operator $\nabla$ in~\eqref{eqn: back-propagation} involves computing  the gradient w.r.t. the model parameters $\mathbf{W}^{t,b-1}$ from the last layer to the first layer backwardly (see details in~\cite{goodfellow2016deep}).  Using the local gradient, the step-$b$ gradient descent refers to updating the local-model parameters as 
\begin{equation}
    \label{eqn: gradient_descent}
    \mathbf{W}^{t,b}_m = \mathbf{W}^{t,b-1}-\lambda \mathbf{G}^{t,b}, \quad b=1, 2, ..., B,
\end{equation}
where $\lambda$ is the step size and ${W}^{t,0}_k={W}^{t}$. After the last mini-batch is processed, worker-$m$ obtains the  local model $\mathbf{W}^{t+1}_m=\mathbf{W}^{t,B}_m$ or the corresponding    local gradient $\mathbf{G}^{t+1}_m=\mathbf{W}^{t,B}_m-\mathbf{W}^{t}$. This completes the effectiveness-encoding process. The uploading of the local model or local gradient ends the current communication round. 

The effectiveness encoding can include the  additional operation of \emph{local model/gradient compression} described as follows. Consider the case of  gradient uploading. A gradient tends to be sparse in the sense that a large number of its elements are much smaller  in magnitude than others. A simple method of gradient compression is to keep a fixed number of elements with the largest magnitudes and set the remaining ones to zeros, thereby substantially reducing the communication overhead~\cite{Deniz2020TSP,Deniz2020TWC}. Consider the case of  gradient uploading. Local models also exhibit sparsity. Parameter (or neurons) pruning can be performed progressively during the process of training  using a suitable metric, for example, variance or  magnitude \cite{Tassiulas2019Arxiv}. A much simpler  method is called~\emph{Dropout} that randomly samples parameters for deletion~\cite{Nader2021Infocom}. Besides reducing  communication overhead, the above model pruning is also effective in avoiding model over-fitting.

\subsubsection{Effectiveness Modulation and Multi-access}
This section aims at overcoming the communication bottleneck in a FL system form the perspective of effectiveness modulation and multi-access. 

\emph{Linear analog modulation} (LNA) supports fast transmission by avoiding the computation-intensive processes of digital modulation, channel encoding and decoding~\cite{Marzetta2006TSP}. Though the  lack of protection by coding limits its application to reliable communication, recently, LNA is gaining popularity in SemCom especially in  fast multimedia transmission~\cite{Takashi2018GLOBECOM} and machine learning~\cite{YQ2019WCL,GX2020TWCBAA,Qin2021JSAC} as human quality-of-experience, machine inference and learning are robust against noise if it is  properly  controlled by, for example, power control and scheduling. In the context of learning, it is even possible to exploit channel noise to accelerate  the learning process by escaping from saddle and local-optimal points~\cite{ZZZ2021Arxiv}. 

\begin{figure*}[t]
        \centering
        \includegraphics[width=0.55\textwidth]{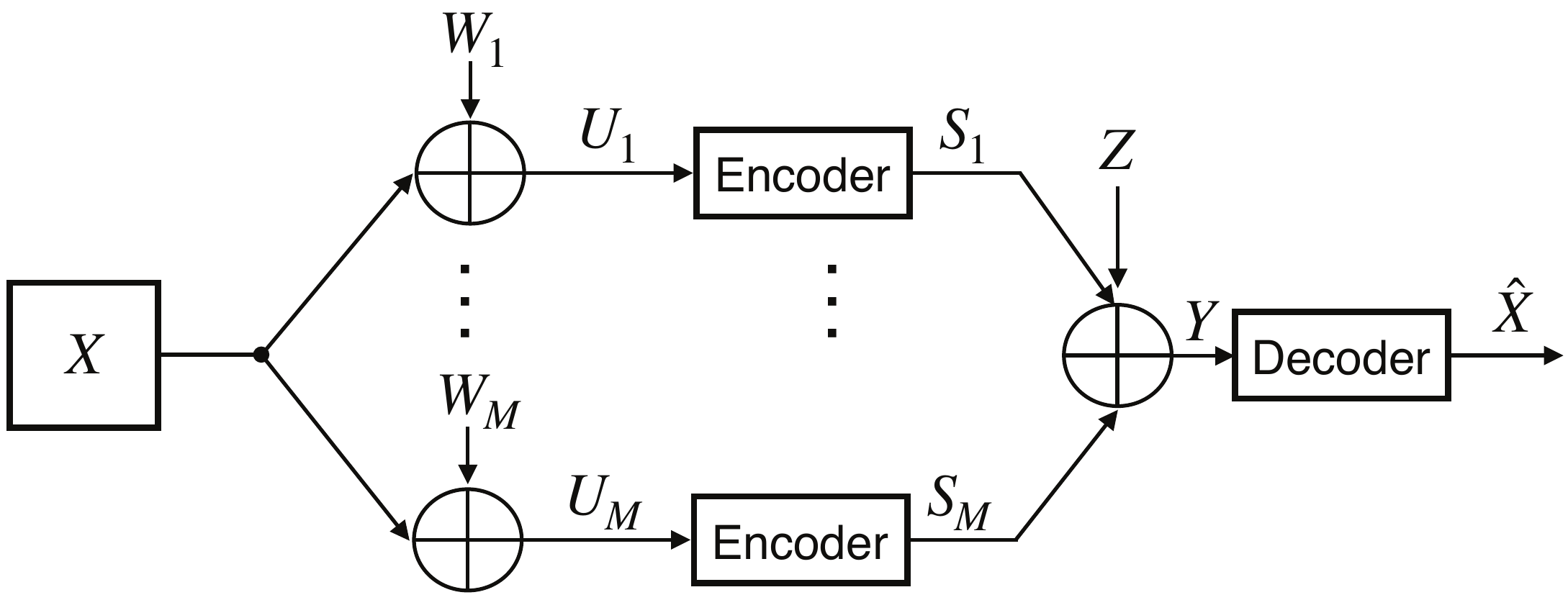}
        \caption{Distributed sensing network.}\label{Fig: UncodedTransmission}
    \end{figure*}

LNA is known to be optimal for the task of distributed sensing in a sensor network as illustrated in  Fig.~\ref{Fig: UncodedTransmission}. The task is to  compute an aggregation function (e.g., averaging) of distributed sensor observations so as to suppress the observation noise. To efficiently carry out the task, a  technique called \emph{over-the-air computing} (AirComp) based on LNA   exploits the waveform superposition property of a wireless channel to perform over-the-air aggregation of simultaneously transmitted sensing data using  LNA \cite{GXAirCompMag}. Let $U_m$ denote a noisy observation at sensor $m$ of a common source $X$: $U_m=X+W_m$ where $W_m$ represents the sensing noise (see Fig.~\ref{Fig: UncodedTransmission}). All observations are transmitted at the same time over Gaussian channels to a server (fusion center) using  uncoded LNA. This results in the following received signal: 
    \begin{align}
    Y=\sum_{m=1}^{M} S_{m} + Z,
    \end{align}
    where $S_{m}$ results from modulating $U_m$  and $Z$ is the Gaussian channel noise. The modulated symbol $S_{m}$ is the scaled version of $U_m$ under the power constraint $\mathbb{E}\left[\left(S_{m}\right)^{2}\right] \leq P_{m}$. The server produces an estimate of  the source $X$, denoted as $\hat{X}$, that minimizes the distortion  $D=\lim _{N \rightarrow \infty} \frac{1}{N} \sum_{n=1}^{N} \mathbb{E}\left[(X[n]-\hat{X}[n])^{2}\right]$, where $n$ represents the symbol index.
    In the presence of channel noise, the server receives the desired average of distributed observations.  It is proved in \cite{GastparUncoded} that in the case of Gaussian sources and noise,  AirComp  achieves  the optimal rate-distortion tradeoff for a large number of sensors, making AirComp an effective multi-access technique for the task. On the other hand, it is sub-optimal for the current task to rely on classic information theoretic encoding that first quantizes the observations into bits and channel encoding the bits. The main reason is the mismatch of the task with the objective of the classic scheme aiming at reliable decoding of data symbols transmitted by sensors. A more vivid interpretation is that AirComp treats interference as a friend rather than a foe.
    
Most recently, AirComp discussed above is applied to realize ``over-the-air aggregation" for fast FL, termed  \emph{over-the-air FL} \cite{GX2020TWCBAA,Deniz2020TWC,Deniz2020TSP}. Given its awareness of the FL algorithm (especially the aggregation operation), AirComp enabled by LNA represents a joint effectiveness design of modulation and multi-access targeting FL. Consider the uploading phase of a communication round and the FL implementation based on local-gradient uploading.  The discussion  can be extended to the other case of local-model uploading straightforwardly. Each worker modulates its   local gradient using LNA and transmits the result  over the same frequency band simultaneously as other workers. By supporting such simultaneous access, the communication latency is reined in to avoid linear scaling with the number of devices, as in conventional orthogonal-access schemes. For over-the-air aggregation it is required to align the magnitudes of the received signals. To this end, each worker performs channel-inversion power control. By synchronizing workers' transmission (by, for example, timing advance) and exploiting  the wave-form superposition property of a multi-access channel, the server receives the desired average of local gradients. Mathematically, each received symbol is  denoted as $y$ and given as
\begin{equation}
    \label{eqn: aircomp_aggr}
    y = \frac{1}{M}\left(\sum_{m=1}^{M} h_{m}p_{m}s_{m} + z\right) = \frac{1}{M}\sum_{m=1}^{M} s_{m} + \frac{z}{M}
\end{equation}
where $M$ is the number of devices and for worker $m$, $s_{m}$ represents a transmitted symbol modulating a single gradient coefficient, $h_{m}$ the  channel gain, and  $p_{m}=\frac{P}{h_{m}}$ channel-inversion power control with $P$ being a given constant, and at last $z$ the channel noise. Over-the-air FL is designed for a broadband system  in \cite{GX2020TWCBAA}  and 
a multi-antenna system in \cite{Liu2020CL,Shi2020TWC}. 

Last, designing over-the-air FL can be based on the proposed SemCom architecture with the layer-coupling approach. In this case, the PAI passed from the Semantic Layer to the Physical Layer is the specifics  of the  aggregation operation (e.g., aggregation weights, selected devices and their uploading frequencies)  in the FL algorithm. 

\subsubsection{Effectiveness Radio Resource Management}
In this subsection, we answer  the effectiveness problem for FL from the perspective of RRM. To overcome the communication bottleneck, RRM should be guided by the principle of allocating more resources to the transmission of data that has a higher importance for the model training, while preventing the unimportant data from occupying  channels. This leads a new class of effectiveness techniques called (data) \emph{importance-aware RRM} \cite{Dongzhu2021TWC, Ren2020TWC,Tao2021TWC, HKB2021TCCN}. 

In a FL system, There exists multiple types of data including  training samples, local gradients, or local models. Regarding training samples for a classifier model, their importance is measured by \emph{data uncertainty}, a popular concept in the area of active learning. 
It is defined as the level that how confident an AI model holds for its prediction to a data sample~\cite{Settles2008NeuralIPS,Dongzhu2021TWC}. Consider a neural-network based  classifier model. A common  metric for measuring the  uncertainty  of a sample $\mathbf{x}$ is the \emph{entropy} of posteriors of  $L$ labels as computed using the model,
\begin{equation}
    \label{eqn: cnn_data_importance}
    u(\mathbf{x}|\mathbf{W}) = \sum_{\ell=1}^{L}\Pr(\ell|\mathbf{x},\mathbf{W})\log{\Pr(\ell|\mathbf{x},\mathbf{W})},
\end{equation}
where $\Pr(\ell|\mathbf{x},\mathbf{W})$ is the posterior of label-$\ell$ given input $\mathbf{x}$ and model parameters $\mathbf{W}$. On the other hand, the importance of gradients can be measure by \emph{gradient divergence}~\cite{Ren2020TWC} or \emph{squared multivariate coefficients of variation} (SMCV)~\cite{Tao2021TWC}. For a local gradient $\mathbf{G}_{m}^{t}$, its gradient divergence is measured by the variance to the global gradient, given by $\big\Vert \frac{N_m}{p_m^{t}\sum_{m=1}^{M}N_m}\text{vec}\left(\mathbf{G}_{m}^{t}\right)  - \text{vec}\left(\mathbb{E}_{m}[\mathbf{G}_{m}^{t}]\right)  \big\Vert^2$ where $p_m^{t}$ is the probability that this local gradient is selected and $\text{vec}(\cdot)$ is the vectorizing operator. The SMCV of an aggregated global gradient vector is given by the sum of means of each entry divided by the sum of variances of each entry with randomness due to channel noise. In addition, the importance of a local model can be measured by its variance to the current global model $\Vert\text{vec}\left(\mathbf{W}_{m}^{t+1}\right)-\text{vec}\left(\mathbf{W}^{t}\right)\Vert$ (see, e.g.,~\cite{Wen2019JCIN} for an overview).

These schemes features both channel and importance awareness and aim at striking a balance between scheduling devices with strong channel for the objective of rate maximization and those with important data for the objective of accelerating model convergence. As a result, the schemes favour devices with either very important data, very strong channel, or satisfactory levels in both aspects. It should be emphasized that in the context of FL, the two objectives mentioned earlier are  not entirely in conflict from the perspective of latency minimization. The former reduces latency per round  but the latter reduces the required number of rounds for model convergence. To minimize the total latency (in second), the above tradeoff should  be optimized. A common  design approach is to derive a DII for implementation using the layer-coupling approach, which  accounts for both the data importance and channel state. Then the  criterion for importance-aware scheduling is simply to maximize the DII.  Consider an edge learning system (e.g., a closed system without the data-privacy issue) directly uploading data from devices to a server for model training. The DII is a linear combination of a channel quality indicator and maximum sample uncertainty of  a local dataset \cite{Dongzhu2021TWC}. Next, consider scheduling for a FL system. Probabilistic scheduling is adopted to avoid a bias of the trained model towards a particular local dataset. To be specific, in each round, each device is scheduled with a given probability. The optimal probability of a device is shown to be proportional to the local-gradient variance and a monotone decreasing function of the communication latency  \cite{Ren2020TWC}. 

Besides scheduling, effectiveness power control schemes have been  also designed for FL to address the issue of \emph{differential privacy}~\cite{Seif2020ISIT,DZ2021JSAC}. By power control, such schemes regulates a sufficiently high  channel-noise level to meet the privacy constraint at the cost of reduced  training accuracy. 

\subsection{Split Inference}
\label{sec:M2MSI}
While the preceding sub-section focuses on model training, the theme of this sub-section is the other facet of machine learning, namely inference  using a trained model. In this area, \emph{split inference} is an emerging paradigm for 5G-and-beyond to offload a large part of the inference task from a mobile device to an edge server hosting a large-scale model~\cite{Zhang2020CM}. The remaining task executed on-device is to extract useful features from raw data for transmission to the server. The task splitting gives the name of split inference. This mitigates the impact of the resource limitation on the device and enriches its capacity via access to a server model, much more powerful and complex than the one that can be afforded as an on-device counterpart. For instance, classifiers in the Google Cloud can recognize thousands of object classes and that in Alibaba Cloud hundreds of waste classes for litter classification. In the remainder of the subsection, we discuss effectiveness coding and communication  separately for the layer-coupling and SplitNet architectures. 
 
\subsubsection{Effectiveness Encoding and Transmission   for Layer-Coupling Approach}
In the  context of split inference, effectiveness encoding refers feature extraction, referring  to the process in which a device encodes high-dimensional raw data into reduced-dimension  \emph{features} or \emph{feature maps}~\cite{goodfellow2016deep}. Features represent information essentially for inference  while  raw data contains a large amount of redundant information (e.g., background objects and noise known as \emph{spatial redundancy} in raw images~\cite{Wang2020NeuralIPS}). Stripping away the redundancy substantially  reduces communication overhead without compromising inference performance.  Our discussion focuses on feature extraction (i.e., effectiveness encoding) while  details on  inference using features (i.e., effectiveness decoding) can be found in a typical  standard machine-learning book (see e.g.,~\cite{goodfellow2016deep}). A classic, simple  technique is  \emph{principal component analysis} (PCA) \cite{Hastie2009Elements}. PCA uses SVD to identify   the most informative low-dimensional linear  subspace (feature space) embedded in a large high-dimensional dataset, called \emph{principle components}. Then projection of a data sample onto the feature space yields its features. Modern feature extraction exploits the powerful representation capability of neural networks and rich training data. Such a feature-extraction model can be implemented using  \emph{multi-layer perceptrons} (MLPs) for a general purpose, CNNs for visual data~\cite{Bajic2021TIP, Ko2018AVSS}, and RNNs for time-series data~\cite{Pagliari2020TC} or leverage  the emerging graph neural networks to improve inference performance with point cloud and non-Euclidean data \cite{JunZhang2021ICASSP}. 

A SemCom system designed for efficient feature transmission is characterized by its \emph{feature-importance awareness}. As widely reported in the deep learning literature, features do not contribute evenly  to inference performance and thus have heterogeneous importance levels ~\cite{Guo2020AAAI}. Available  importance measures include \emph{divergence}  for data statistical models (e.g., discriminant  gains of specific feature dimensions)  \cite{Saon00NIPS} and other classification-loss related metrics for DNN models~\cite{Guo2020AAAI}. 
Consider an importance-aware SemCom system designed using the layer-coupling approach. The  CRI passed to the discussed effectiveness encoder controls the number of features to extract. Given the number, features are selected based on their importance levels (DII) to be transmitted in the radio-access layers~\cite{QiaoLan2021TechReport}. There exist numerous algorithms for feature pruning for neural networks (see e.g.,~\cite{Zhuang2018NIPS}). Some design supports channel adaptation of encoding under given requirements on  latency and inference performance~\cite{ShengZhou2019Infocom}. The DII is also passed to the layers for importance aware quantization  (e.g., more important features have higher resolutions) and RRM (e.g., more bandwidth/time-slots for more important features) \cite{Bajic2021TIP, Park2021Sensors,Choi2018CVPR}. Moreover, the DII also determines the transmission sequence (i.e., more important features are transmitted first) so that transmission can be stopped earlier under a inference-uncertainty requirement~\cite{QiaoLan2021TechReport}. On the other hand, PAI providing some information on the effectiveness encoder (e.g., its type or architecture) can be useful for the choice of a matched  classifier model at the server and thus passed to the latter.

\subsubsection{Effectiveness Encoding and Transmission for SplitNet}
\label{subsec: ae_based_splitnet_for_inference}

Consider the implementation of split inference on the SplitNet architecture in Fig.~\ref{Fig: JSCC} with semantic encoder/decoder replaced by their effectiveness counterparts targeting the task of inference. The function of the effectiveness  encoder is to extract feature based on designs  discussed in the preceding sub-section. On the other hand, the effectiveness  decoder is a neural network performing inference. The popular approach of designing the pair of channel encoder/decoder  is to use AE~\cite{goodfellow2016deep}. An AE comprises an encoder and an decoder. Generally, the AE's encoder compresses high-dimensional  inputs to reduced-dimension  outputs; using them as inputs, the decoder attempts to reconstruct the encoder's inputs. In a split-inference system, the two AE components interface with a wireless channel (see Fig.~\ref{Fig: JSCC}). Then the AE based channel encoder directly maps features  to analog modulated channel symbols and the channel decoder decode received symbols into features as input to the subsequent effectiveness  decoder to generate inference results~\cite{Gunduz2021JSAC}.  The design of semantic and channel encoders are under two constraints. First, given $B$ complex channel symbols, the number of extracted features  (real scalars) should be $2B$. Second, a normalization layer is required in the channel encoder such that channel symbols can satisfy \emph{transmit power constraints}. The end-to-end training of the encoders/decoders in SplitNet is difficult to a large number of layers in the combined global model and also channel hostility (i.e., fading and noise) embedded in it. This difficulty is overcome by training the two AE components separately from the semantic encoder/decoder. Specifically, the effectiveness  encoder and decoder are pre-trained in advance since they are independent of the channel and remain unchanged even if the channel statistics vary \cite{Gunduz2021JSAC}. On the other hand, the AE based channel encoder and decoder can be quickly retrained using transfer learning as the radio-propagation environment changes~\cite{Lee2019Access}. This provides the components capability to cope with channel noise.  As a final step, an end-to-end training of all neural is conducted so that they can be further adjusted to achieve optimal end-to-end inference performance in the presence of channel hostility. 

Split inference involves a \emph{computation-communication tradeoff}, described as follows. The effectiveness  encoder and decoder in the SplitNet architecture (see Fig.~\ref{Fig: JSCC}) can be generated by splitting a single AI model (i.e., a neural network) into two parts with unequal numbers of layers. Shifting the \emph{split point} to the left results in  simpler  on-device effectiveness  encoding and and higher complexity for effectiveness decoding at the server, and vice versa. Intuitively, as the device is  resource constrained, it is desirable to push the split point as close to the input layer of the AI model as possible. The intuition is correct from the perspective of computational load but overlooks the other perspective  of communication overhead. Specifically, in a large class of popular AI models in practice, the size of features output by ``shallow"  feature-extraction layers is large and can be even much larger  than that of  raw data at the input, which is known as  ``data amplification effect''~\cite{Li2018IPADS}.  Consequently, a shallow split point may results in unacceptably large  communication overhead and  energy consumption, defeating the original purpose of split inference. This motivates researchers to adjust  the split point with the aim of optimizing  the communication-and-computation tradeoff \cite{XuChen2020TWC,Bennis2021arxiv}. Relevant algorithms rely on profiling  the operational  statistics of individual model layers including feature size, latency, energy consumption,  and required memory size. Then the profiles are applied to design algorithms for adapting the split point to the time-varying communication rate under latency requirements and devices' resource constraints. 

Last, the required connectivity type for split inference depends on the specific application. For the  family of mission critical applications (e.g.,  finance, auto-driving,  and automated factories),  URLLC connectivity is required \cite{Han2021WC}). For instance, remote inference for autonomous driving is expected to have  $1$ ms latency and near $100\%$ reliability in communication~\cite{TR22891_New,Campolo2017WC}. Other applications are not latency sensitive but require the close-to-human machine  vision (i.e., recognition of hundreds of object classes for a high-end surveillance camera), eMBB will be needed to transmit high-dimensional features extracted  from high-definition images.

\subsection{Distributed Consensus}\label{sec:M2MDD}

Distributed consensus refers to the process that agents in a distributed network act together to reach  an agreement by message exchange. A typical algorithm  involves each agent interactively updates its own state based on received information on peers' states \cite{Ricker2007TAC}. When there are many agents, the convergence could be slow and as a result the iterative process could incur excessive communication overhead, for example, in the specific scenarios of  vehicle platooning~\cite{Hult2016spm} and blockchains~\cite{Petar2019IoT-J}. To address this issue, the criterion for designing SemCom for efficient distributed consensus is to reduce the overhead without significantly decreasing the convergence speed. The key component is the design of effectiveness encoding that is aware  of the algorithm and its objective and based on  the knowledge, extracts and transmits semantic information from an agent's state to others. In the remainder of this subsection, we introduce  two representative scenarios of distributed consensus, namely vehicles platooning and blockchains, and discuss matching effectiveness coding techniques. 

\subsubsection{Vehicle Platooning}
Vehicle platooning  is a  high-way automatic transportation method for driving a cluster of  connected vehicles in a formation  (e.g., a line) to achieve higher road capacity and greater fuel economy~\cite{Winnie2016TITS}. This requires the member vehicles to brake and accelerate together based on the system state, which represents the consensus. Maintaining the system state requires vehicles  to continuously share and update their local states e.g.,  vehicle parameters (e.g., positions, accelerations and velocities), sensing data (e.g., traffic lights, pedestrians, obstacles, road conditions, and LIDAR imaged point cloud),  and even individual auto-pilot models. Transmitting all the raw state data is impractical. For instance, an typical  autonomous vehicle collects from its sensors up to several gigabytes of  data per seconds. Thus, it is essential  to design effectiveness encoding to extract from the raw state  data the information essential for convergence to a consensus. To better understand the principle of a vehicle-platooning algorithm, consider a simple scenario of driving a platoon along a straight high-way in a line formation. In this case, effectiveness encoder  of each vehicle, say vehicle $m$,   outputs its distance to its predecessor, $s_m$, and its own speed,  $v_m$, which defines the local state, while its control variable is its acceleration, $a_m$. The local states are assumed to be exchanged continuously between vehicles over wireless links. Let $r_m(\tau) $ denote  the cost  at time $\tau$ for front situation (i.e., the relationship between vehicle-$m$ and its predecessor, vehicle-$(m-1)$). Typically,  $r_{m}(\tau)$ accounts for  all or some  of the following aspects, namely safety cost, efficiency cost and comfort cost, each of which can be defined as function of the states of vehicle $m$ and those of its neighbours (see examples in \cite{Winnie2016TITS,Monica2017TITS}). Hence $r_{m+1}(\tau) $ represents the behind situation of vehicle $m$. Then the local control problem at the vehicle over  a duration  $T$ and with the objective of behind-and-front cost minimization  can be formulated as ~\cite{Winnie2016TITS}
\begin{equation}
    \label{eqn: platoon_local_obj}
    \min_{\sf a_m} \int_{0}^{T} r_m(\tau) + r_{m+1}(\tau) \mathrm{d} \tau,
\end{equation}
where $\tau=0$ denotes the current time instance. Iteratively solving the problem, applying the computed acceleration, and broadcasting the local state by  all  vehicles will eventually reach their  consensus  on the platoon's optimal  speed and inter-vehicle separations gaps. There exists a tradeoff between: 1) the complexity of effectiveness encoding and the amount of its output information (that determines communication overhead), and 2) the sophistication of the platooning algorithm. For example, a vehicle's predicted trajectories can be shared with others in the platoon, requiring  effectiveness encoders to compress the trajectories. Most recently, deep learning have been adopted to  empower platooning.  Essentially, CNN-based effectiveness encoders are designed  to intelligently extract information from real-time videos captured by on-board cameras, such as traffic lights, lanes and obstacles~\cite{Siddique2019IJDSN}. Exchanging such sensing data and use them for consensus on complex manoeuvres give  the platoon collective  intelligence for auto-driving.

Other  SemCom techniques  have been extensively studied in the literature. First,  URLLC connectivity is required in this mission-critical application  to avoid  collisions \cite{JPark2021arxiv}. In terms of latency for vehicle platooning, it should be measured and minimized in terms of \emph{information latency} rather than the conventional over-the-air latency as  the former  directly relates  to coordinated  control performance~\cite{Niu2020WC}. To overcome the limit of radio resources, its effectiveness allocation for vehicular platooning  should be \emph{importance aware} by identifying critical and less critical information in vehicles' state data, allowing them to be compressed accordingly to the specific driving algorithm ~\cite{Hult2016spm}. On the other hand, it is proposed in \cite{JPark2021arxiv} that effectiveness RRM and multi-access should also have \emph{situation awareness} and be optimized for a specific  vehicular network topology represented using a graph. Based on the principles, SemCom techniques are designed based multi-agent RL to integrate  the  operations in the semantic layer and physical layer (e.g.,  transmission stopping), thereby reducing the intensity of communication.

\subsubsection{Blockchains} 

A \emph{blockchain} is a growing  chain of \emph{blocks}, each of which  contains a time stamp (when the block was published), transaction data  of the blockchain, and cryptographic information of the previous block. In this way, the chain is robust against any alteration of the   transaction data by an individual block  as it requires changes on subsequent blocks too. As they  can  implement  public distributed ledgers, blockchains find a broad range of applications ranging from crytocurrencies to gaming to financial services~\cite{Li2021TPDS}. 
In a distributed network  containing a blockchain, the devices are \emph{nodes} within that blockchain. Nodes can  propose changes to the blockchain by submitting \emph{transactions} via broadcasting to all other nodes. One distinctive   feature of the blockchain protocols is that a transaction {should} be broadcast to all member nodes {in order to reach} consensus. For this reason, {transaction approval} relies  on frequent communication to exchange information {and reach a consensus across} a large number of nodes. This motivates the design of effectiveness encoding for blockchains.

{Let us take the example of application of} blockchains to building construction~\cite{XUE2020AutoCons}. A large-scale  project involves a large team and many contractors/sub-contractors that perform  distributed field works. A  blockchain can be  used as as a secure distributed ledger to facilitate cooperation  and  ensure construction quality. Based on this platform, the physical and functional features of building components are stored in blocks and validated by all parties. During the construction, a  change made on  a particular  component (e.g., a  new design   or construction progress) by a stakeholder will trigger  updating of all nodes in the blockchain. For this to happen, the change in question has to be submitted as a  updating proposal (i.e., a transaction) and  approved by other nodes upon validation before it is made on the blockchain. The detailed digital format of a transaction depends on on the choice of data model structure for that blockchain, such as the Industry Foundation Classes schema for civil engineering, and the transaction algorithm. One design of effectiveness encoding for communicating transactions is based on transmitting \emph{differential states}, called \emph{semantic difference transaction} (SDT) in \cite{XUE2020AutoCons}. Specifically, the SDT-based encoder compares the objects in the new schema  with the validated ones recorded in the blockchain, aiming to identify  the  objects that require updating. Then the encoder generates  the required changes of only the identified objects, which is broadcast to all other nodes for validation. {Compared the case in which the whole schema is broadcast,} SDT can substantially reduce the communication overhead, especially given that the changes are usually minor.

Moreover, there exists fault-tolerant consensus protocols for further reduction of the communication overhead via {effectiveness-based} resource allocation. In the notion of \emph{practical Byzantine fault tolerance} (PBFT) consensus, an {effectiveness-based} resource allocation mechanism groups nodes into layers and only executes inner-layer communication of transactions for convergence of consensus with a given security threshold~\cite{Li2021TPDS}.

\subsection{Machine-vision Cameras}\label{sec:M2MDS}

Machine vision cameras, which  are connected by IoT and relies on servers for sensing data analysis, are capable of identifying interested labels in recorded images and videos  such as time, location and objects~\cite{Ren2018Network}. They are commonly used as standalone cameras or as a surveillance network deployed in homes, factories, and cities  to detect  human gestures and activities~\cite{Chen2017TCSVT}, for security management~\cite{Sultana2019Access}, or identify defective products on a production line~\cite{Ota2018TII}. At a larger scale, machine-vision cameras are merged into aerial and space sensing networks to form a universal network~\cite{Skinnemoen2014AERST}. The communication bottleneck of  a machine-vision camera network arises from  large-size raw data generated by each camera and the enormous camera population (e.g., millions of connected surveillance cameras in a metropolitan city)~\cite{Kim2019Mobicom}. A single  frame in $1080$P videos consists of two million pixels while there can be up to $60$ frames per second, generating  data at a  rate of  $100$Mbps~\cite{Gao2013ISCAS}. In the sequel, we discuss effectiveness encoding and RRM for efficient SemCom in such a network. 

One  key feature of effectiveness encoding is to detect \emph{regions of interests} (ROIs) in  a set of visual data that  contain interesting  labels and thereby facilitating trimming of videos or images for efficient streaming  to edge or cloud servers for  analysis \cite{Ren2018Network}. A CNN model is  commonly deployed  as an effectiveness encoder  to detect ROIs. Specifically, consider a set of $K$ frames (or images), denoted as  $\{\mathbf{I}_k\}_{k=1}^{K}$, each of which, say frame $k$,  comprises $R$ regions, denoted as $\{\mathbf{I}_{k,r}\}_{r=1}^{R}$. A lightweight on-camera CNN detector  scans  each region of every frame to search for interesting objects. For  frame $k$, the indices of spatial ROI  will be grouped into the index set  $\mathcal{F}_{\sf sR}(k)$ defined as 
$\mathcal{F}_{\sf sR}(k) = \{ r \ \vert \ \text{objects captured in $\mathbf{I}_{k,r}$}\}$. Then the number of spatial ROI is $\vert\mathcal{F}_{\sf sR}(k)\vert$.
In  the \emph{temporal} dimension, frames comprising interesting objects are then  included into the index set  $\mathcal{F}_{\sf tR}$ defined as $\mathcal{F}_{\sf tR}=\{\mathbf{I}_k \ \vert \ \vert\mathcal{F}_{\sf sR}(k)\vert>0\}$. Consider  security management as an example \cite{Sultana2019Access}. A region of a frame containing dangerous objects such as  knives and guns will be tagged as a spatial  ROI. The temporal ROIs sets $\mathcal{F}_{\sf tR}$ are then encoded and transmitted to servers for further analysis while those frames not in $\mathcal{F}_{\sf tR}$ can be coarsely compressed or even discarded.

While conventional RRM  schemes  deliver  video bits indiscriminately,  effectiveness designs targeting machine-vision cameras differentiate the importance level of sensing data given their relevance to ROIs, which can be used as DII in the layer-coupling approach. In terms of quantization, more bits can be  allocated to high-resolution quntization of the pixel regions in $\mathcal{F}_{\sf sR}(k)$ and fewer bits to quantizing  background pixels~\cite{Ren2018Network}. In the presence of multiple cameras, the quality of contents from  each camera should be assessed in terms of  how critical they are for executing a given  task. Cameras capturing critical ROIs should be given a  higher priority in RRM~\cite{Chen2017TCSVT}. Besides ROI detection, it is possible for cameras with increasing computation capacity to perform part of data analysis and extract features  from multimedia sensing data using a DNN model 
and a  knowledge base~\cite{Kot2020TIP}. Last, it should be mentioned that given  the above operations, IoT connected computer-vision  cameras can be implemented using either the layer-coupling or the SplitNet approaches, discussed earlier.

\section{KG based Semantic Communications}\label{Section:KG}
A KG is composed of the representations of many entities in a semantic space and the relations among them. KGs have become a powerful tool for interpretation and inference over facts~\cite{Inference,RuleReasoning}. Many massive KGs have been constructed including  Wikidata \cite{Wiki}, Google KG \cite{Google}, WorldNet~\cite{WorldNet}, Cyc \cite{Cyc}, YAGO~\cite{YAGO}. They have formed  the foundation for Internet and a knowledge base for understanding 
how the world works. In particular, large-scale  KGs are used by search engines such as Google, chatbot services such as Apple's Siri, and social networks such as Facebook. In this section, we introduce a paradigm of SemCom featuring the use of KGs as a tool to improve communication efficiency and effectiveness. In this context, the key function of a KG is to provide a  semantic representation of information such that semantic  encoding is not only efficient but also robust against communication errors. For H2H communication in the presence of errors, a KG based decoder can correct the errors by decoding the received erroneous message  as a correct one with largest similarity on the graph \cite{ERNIE1,ERNIE2}. For H2M symbiosis, a KG can function as a set of human behavior rules to exclude unreasonable results due to faulty sensing results \cite{RuleBased1,RuleBased2,RuleBased3}. Furthermore, for M2M SemCom, KGs are useful in knowledge sharing between different types of  machines and thereby serving as machine interfaces in heterogeneous networks. In the remainder of the section, we will provide a preliminary on KG theory and then discuss KG based SemCom techniques, applications, and architectures.

\subsection{Preliminary on KG Theory}

\begin{figure*}[t]
    \centering
    \subfloat[Knowledge represented by factual triples.]{\includegraphics[height=4.4cm]{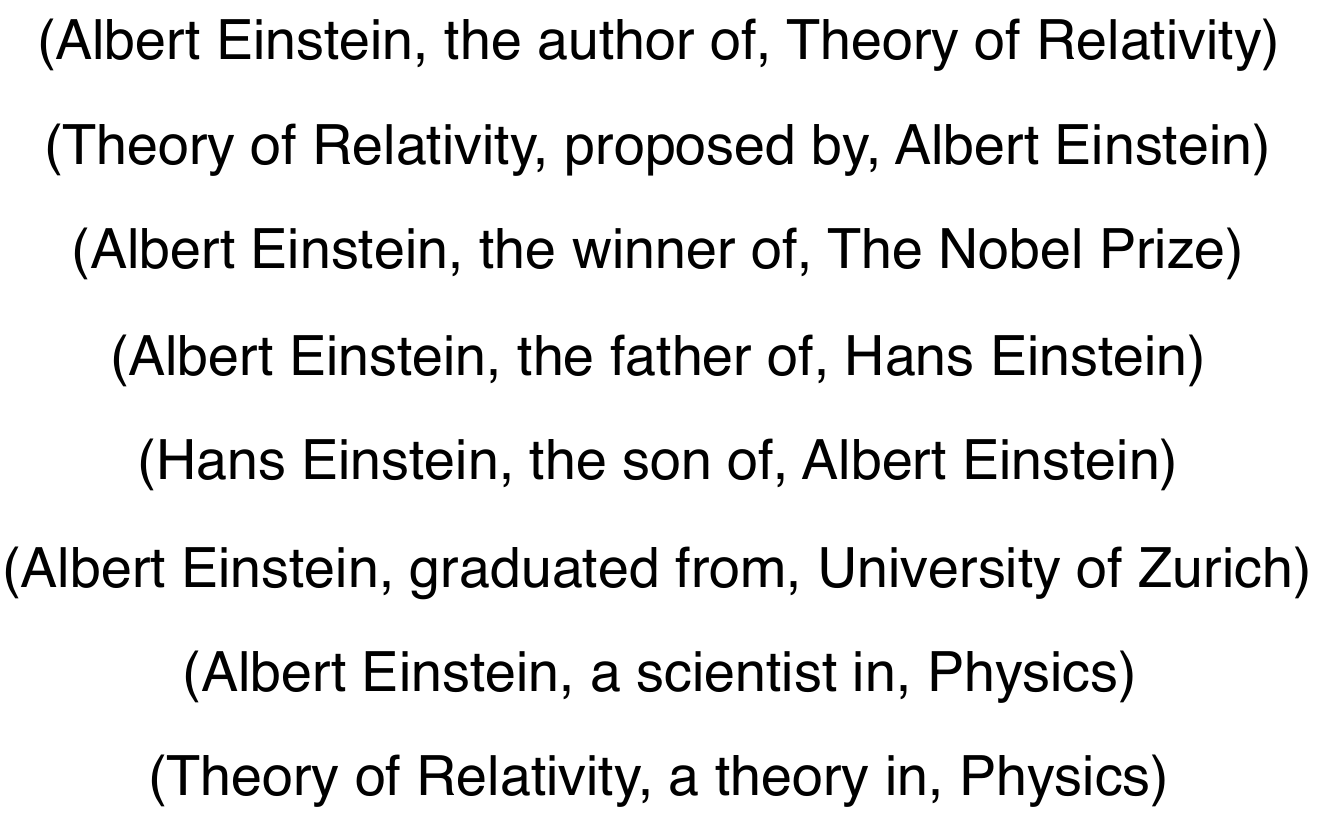}}
    \subfloat[Entities and relations represented by a KG.]{\includegraphics[height=4.7cm]{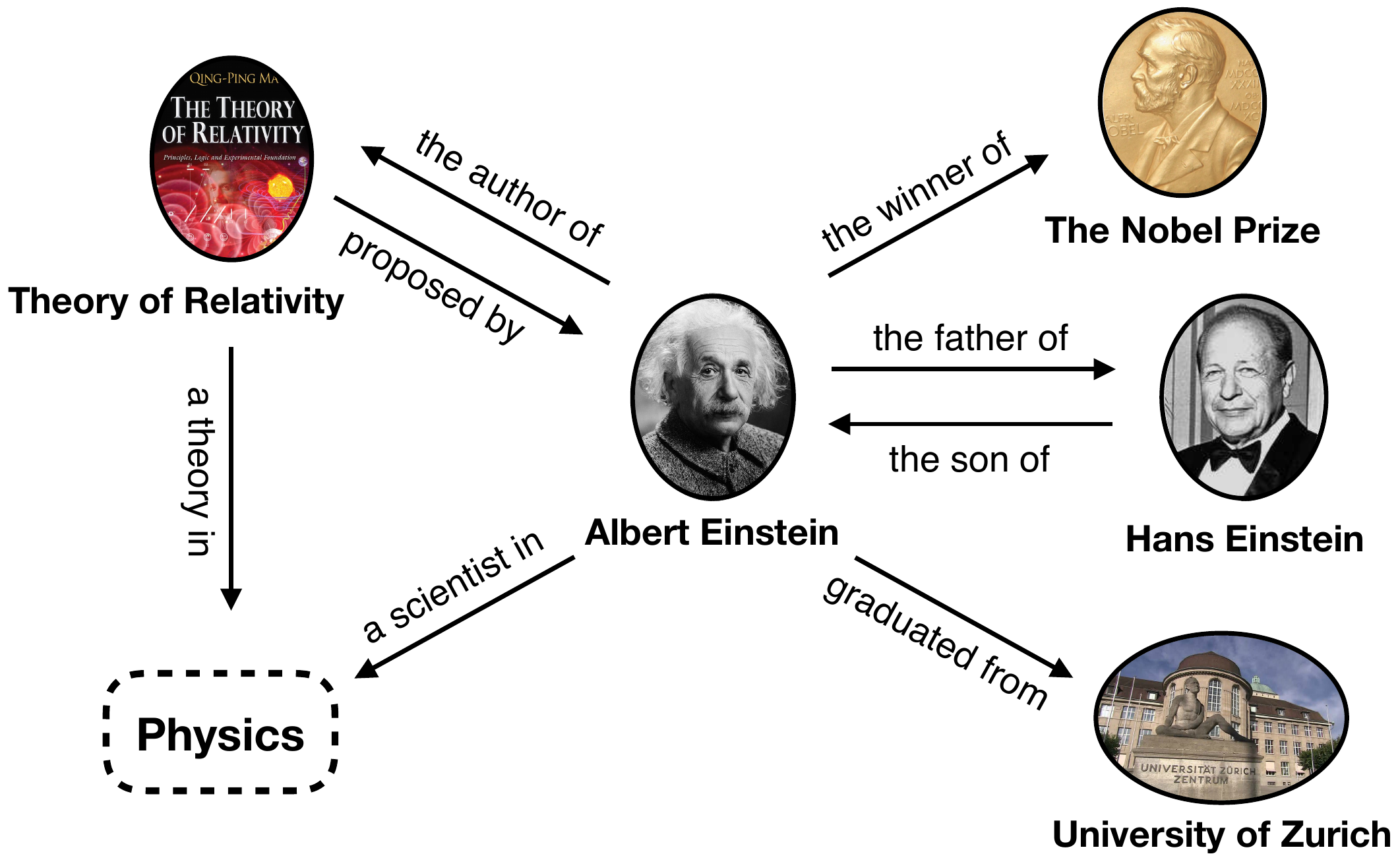}}
    \caption{A example of KG from \cite{knowledgeSurvey}.} \label{KGofEinstein}
\end{figure*}

KG refers to the broad area of graph representation of knowledge without a unified definition \cite{knowledgeSurvey, KnowledgeBaseGraph}. For concrete discussion, we consider the definition introduced in \cite{knowledgeSurvey} where nodes are nouns related to real-world objects/names/concepts and edges specify their relations. One example  is illustrated in Fig. \ref{KGofEinstein}. A fact, a basic element of knowledge, can be represented by a so-called factual triple (head node, relation, tail node) or mathematically $(\mathbf{h},\mathbf{r},\mathbf{t})$, e.g., (Albert Einstein, Graduated From, University of Zurich).  A node (e.g., $\mathbf{h}$  or $\mathbf{t})$) is a vector, say a $L$-dimensional vector,  storing relevant information, creating a $L$-dimensional semantic space. For instance, if Einstein is the head node $\mathbf{h}$, the tail node $\mathbf{t}$ can be ``Theory of Relativity", ``The Nobel Prize", ``University of Zurich", ``Hans Einstein" (his son), and so on. The relation $\mathbf{r}$ is either a  vector if the mapping is \emph{distance based}, i.e., $\mathbf{h}+\mathbf{r} = \mathbf{t}$, or  a matrix (re-denote $\mathbf{r}$ as $\mathbf{M}_r$)  if the mapping \emph{semantic-similarity based}, i.e., $\mathbf{h}^T\mathbf{M}_r = \mathbf{t}^T$. The knowledge relevant to an object, such as a human, is potentially infinite. A KG reduces the infinite knowledge to a finite-dimensional semantic space to enable  practical knowledge processing and transmission.

One potential issue that can arise  from the finite dimensionality of a KG is that a fact involving two nodes, $\mathbf{h}$ and $\mathbf{t}$,  is still plausible even if it is not captured by the graph due to a missing edge/relation connecting the nodes.  The issue can be addressed by introducing a scoring function measuring plausibility. Two typical functions, namely the distance based and a semantic-similarity based function,  are given as follows~\cite{Distance2,KnowledgeBaseGraph} 
\begin{align}
    f_{r}(h, t)=\left\|\mathbf{M}_{r, 1} \mathbf{h}-\mathbf{M}_{r, 2} \mathbf{t}\right\|,\quad f_{r}(h, t)=\|\mathbf{h}+\mathbf{r}-\mathbf{t}\|,
\end{align}
where $\mathbf{M}_{r, 1}$ and $\mathbf{M}_{r, 2}$ are two relation matrices (edges) of  the KG. 

Provisioned with sets of valid and invalid facts, a KG can be constructed using either the rule-based  or the data-driven approach. Either approach requires the definition  of a suitable loss function. One typical choice is the margin-based function given as \cite{marginBasedLoss}
\begin{align}\label{Loss1}
    \sum_{(h, r, t) \in \mathcal{F}} \sum_{\left(h^{\prime}, r, t^{\prime}\right) \in \mathcal{F}^{\prime}} \max \left(0, f_{r}(h, t)+\gamma-f_{r}\left(h^{\prime}, t^{\prime}\right)\right),
\end{align}
where $\mathcal{F}$ and $\mathcal{F}^\prime$ represent  the set of valid and invalid  triples,  respectively, and $\gamma$  \eqref{Loss1} is a given  margin. Other available designs include logistic based and cross-entropy based functions  \cite{crossEntropyLoss1,crossEntropyLoss2,Qin2021DeepSC}.

KGs are useful for training  AI models  especially those  with semantic requirements such as  linguistic applications and involving human-machine interaction.  The structured knowledge in a KG reduces the search complexity in training and helps improve the accuracy of a trained model. Success has been demonstrated in the areas of question answering \cite{questionAnswering1,questionAnswering2}, virtual assistants~\cite{virtualAssistants1,virtualAssistants2}, dialogue~\cite{Dialogue},   and recommendation systems~\cite{Recommendation1,Recommendation2,Recommendation3}. Some KG based techniques and their use in  SemCom are discussed  in the sequel.

\begin{figure*}[t]
    \centering
    \includegraphics[width=0.8\textwidth]{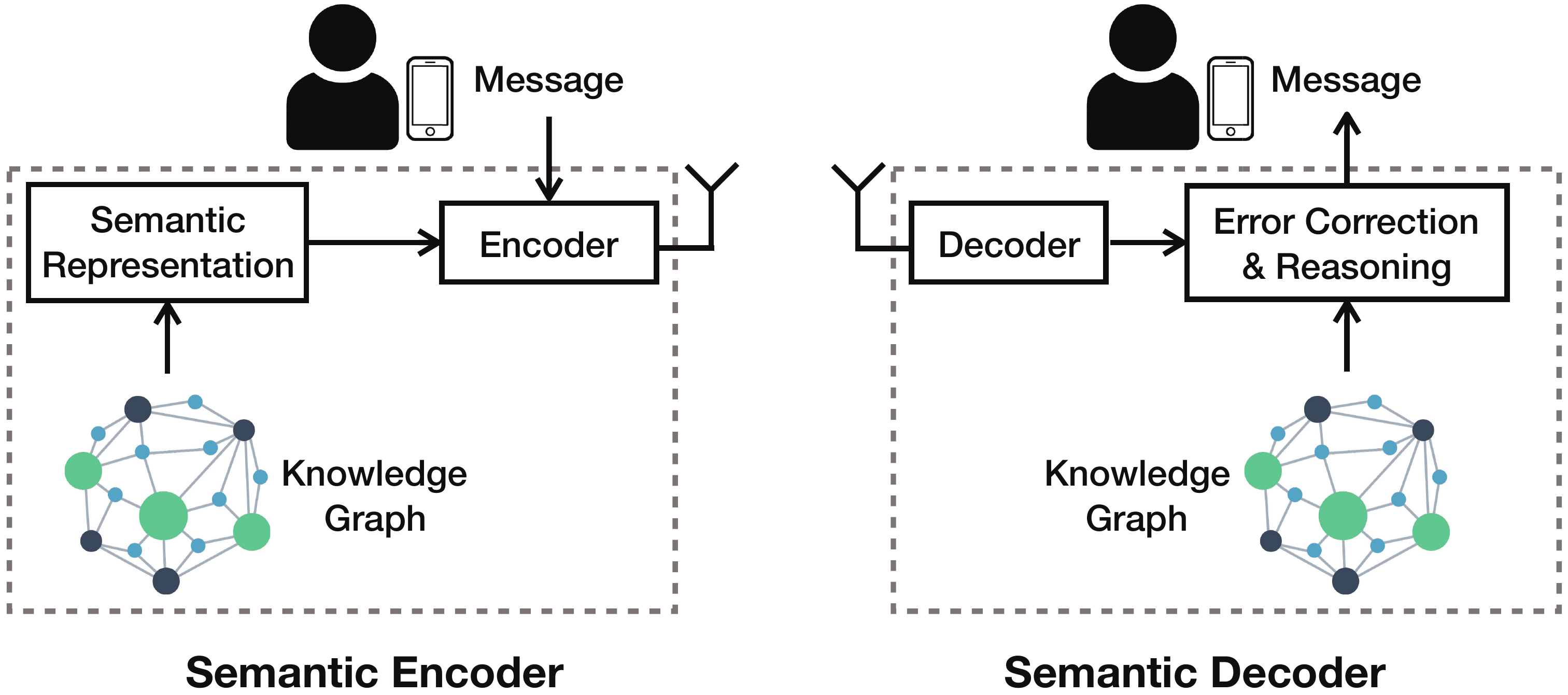}
    \caption{KG based H2M semantic communications.}\label{Fig:KGHH}
\end{figure*}

\subsection{KG  based  H2H SemCom}

For H2H SemCom, a KG representing  knowledge on the background of the parties or the domain of  their conversation can be injected into a semantic encoder to boost SemCom efficiency and robustness  \cite{ERNIE1,ERNIE2,SEED}. As a concrete  example, we discuss the use of the design presented in  \cite{ERNIE2} to encode a simple sentence  \emph{``Albert Einstein won the Nobel Prize for physics in 1921''}. The most important  component of the KG assisted encoder is  a knowledge encoder. Consider input tokens representing individual words in the input sentence/message. Define an \emph{entity embedding} as a node of the KG to which some  tokens of the input sentence can be mapped. For instance, the words/tokens  ``Albert'' and ``Einstein'' can be mapped to the node ``Albert Einstein''   of the KG in Fig. \ref{KGofEinstein} and hence share the same entity embedding. Similarly, the words namely  ``Nobel", ``Prize", and ``physics" are mapped to corresponding nodes of the KG. The distinctive feature of a knowledge encoder is to fuse the tokens of the original message with their entity embeddings to generate  output  tokens  as well as their  embeddings targeting a specific task. The output tokens carry  not only the   information  of the input tokens but also that of others mapping to the same entities. For example, an input token  ``Albert'' would generate the output  ``Albert'' and ``Einstein''. 
 The  encoder is made of stacked aggregators, each further consisting of two multi-head self-attention modules. Note that each such module is designed  to concatenate multiple self-attention modules, each of which  relates  different positions of the input single sequence to compute a representation of the sequence \cite{ERNIE1}. The use of a knowledge encoder at a semantic receiver can exploit a KG to correct inaccuracy  in the semantic meaning  and fill some missing tokens of a received  message as caused by channel errors during the transmission.

\subsection{KG based H2M SemCom}
For H2M SemCom, a KG helps a  machine to understand the current context and the semantic information embedded in the received message   from human beings and react intelligently \cite{humanMachineRequirement,ContextKnowledge}. To be more specific, training a model underpinning  the machine using  structured knowledge gives it the  ability  to recognize the entities embedded in the received messages  and their  relations to other entities, which helps the generation of logical  reactions. Such an approach has found applications in  question answering, dialogue and recommendation systems \cite{Recommendation3,Dialogue}. In the area of robotics, imitation learning has been designed based on  a knowledge-driven approach where the robotic assistants imitate human by inferring semantic meanings in the observed human actions \cite{Imitation1,Imitation2}. Furthermore, in human sensing applications, KGs can be used to define a set of human behavior rules. 

For concrete discussion, the remainder of the sub-section focuses on the use of KG in utterance generation, a basic topic in conversational AI assistants. The task is to 
generate relevant utterances (sentences or phrases) from a knowledge base. In this area, a \emph{long-short-term memory} (LSTM) network is widely used. It refers to  a specific RNN with long-term memory for the important and consistent information while short-term memory for the unimportant information. The effectiveness of LSTM networks has been proven in enabling a machine to generate   utterances based on the received message, the knowledge base as well as its dialogue history with the human partner \cite{questionAnswering1,Dialogue}. On the other hand, feeding the DBpedia KG corresponding to the Wikipedia database into a LSTM network makes it capable of interpreting questions and producing reasonable answers. There exist other  designs. In \cite{questionAnswering2}, a KG is provided to a CNN to extract  semantic features of an input question for subsequent answer searching. On the other hand, a knowledge-driven multi-model dialogue system designed  in \cite{virtualAssistants1}  is capable of gesture recognition, image/video recognition,  and speech recognition, providing multi-model human-like abilities to virtual assistants. Futhermore, 
KGs can also render the operations of recommendation systems more explainable \cite{Recommendation1,Recommendation2,Recommendation3}.

\begin{figure*}[t]
    \centering
    \includegraphics[width=0.75\textwidth]{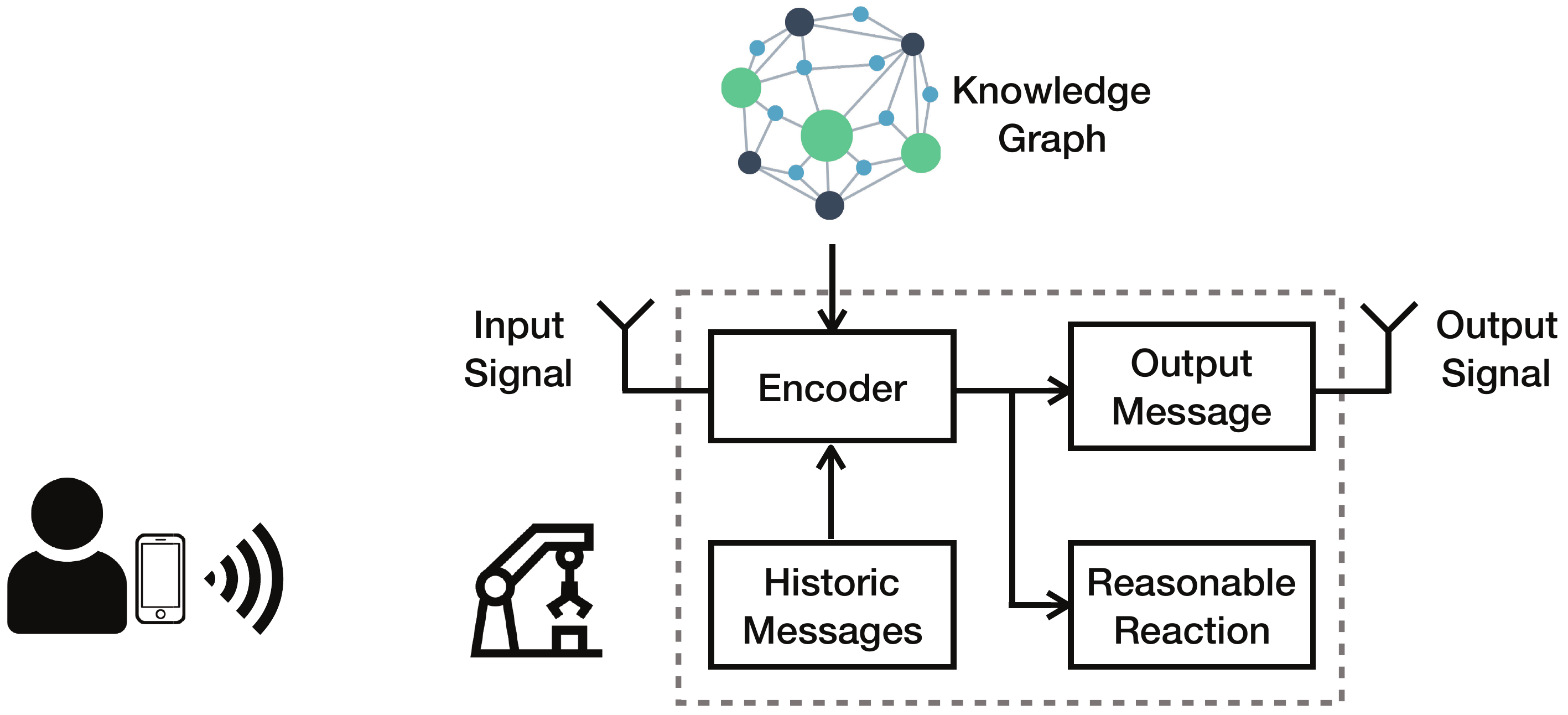}
    \caption{KG based H2M semantic communications.}\label{Fig:KGHM}
\end{figure*}

\subsection{KG and  M2M SemCom}
KGs are related to M2M SemCom in several ways. First, KGs can provide a platform for implementing large-scale IoT networks such as smart cities, logistics networks,  and vehicular networks. Consider a vehicular network as an example. A large-scale dynamic KG can be constructed and periodically updated to represent the states of connected   vehicles (e.g., locations,  velocities, acceleration, routes, and destinations) and their relationships (e.g., chances of collision and whether they form platoons) \cite{KGPlatoonM2M}. Such a KG paves a foundation for facilitating vehicle-to-vehicle communication (e.g., exchange of state information) to avoid accidents or facilitate platooning as well as a platform for traffic management and operating ride-sharing or car-hailing services.  Second, KGs provide a tool for managing SemCom or other types of  networks to facilitate resource allocation, work-flow recommendation, and service selection \cite{Manage1,Manage2,Manage3}. The machine intelligence needed for efficient network management can be powered by structured knowledge embedded in a network KG. Such a KG can be constructed to contain the network topology,  requirements of different applications, expert knowledge from community data, product documents, engineer experience reports, user feedback, etc. Third, an M2M SemCom system can be deployed to support the extension and updating  of a KG. In particular, SemCom between  a large number of edge devices enables distributed knowledge extraction, storage, and fusion \cite{Distributed1,Distributed2}. To this end, each device obtains up-to-date local knowledge via interaction  with its environment and uploading the real-time knowledge to servers for fusion and updating the global KG \cite{Fusion1,Fusion2,Fusion3,Fusion4}. The efficient knowledge transmission can rely on some efficient SemCom technique discussed in preceding sections (e.g., importance-aware transmission). Last, KGs can provide a tool for enabling inter-operability which is necessary for M2M SemCom in cross-domain applications, where the knowledge and information of devices of heterogeneous types have to be shared or aggregated \cite{M2M1,M2M2,SensorWeb}. One particular architecture for such a purpose is proposed  in \cite{M2M2}. It uses  a server as a semantic core to exchange  the messages  sent by  heterogeneous devices by serving as both a relay and a semantic encoder that translates a message from one machine language into another.  

\subsection{KG based SemCom Architectures}

First, consider the SplitNet architecture discussed in Section \ref{sec:jscc}. The use of KGs in training AE and auto-decoders has been demonstrated to improve their capabilities to decode the correct semantic meanings from the received messages despite their distortion by communication channels \cite{ERNIE2}. A related but different approach is proposed in 
\cite{ERNIE1}, where combining source information and its corresponding representation in a KG as inputs to the AE is shown to enhance the SemCom performance. Next, an architecture featuring KG server assisted SemCom is presented in \cite{Shi2020FromSC}. The server located at the network edge relies on a KG to interpret the semantic  meaning of the messages sent by a source device, efficiently encode/translate the messages, and then relay the results to the destination device. As a comprehensive KG can have an enormous size, its storage and inference complexity  far exceeds the  capacities of devices.  Offloading the KG to a  server overcomes the limitations of devices to exploit the KG for reducing the SemCom overhead.

\section{Towards 6G Semantic Communication}\label{Section:conclusion}
While the 5th generation of mobile networks are being rolled out around the world, the global research on 6G has started with an accelerating pace so that the  technologies can  be ready for commercialization in 2030 \cite{Tataria2021-6G,Mahmood2020,Saad20206GVision}. Compared with preceding generations, 6G will achieve limitless connectivity that will scale up  IoT to become  \emph{Internet-of-Everything} (IoE)  and revolutionize networks  by connecting human beings to intelligent machines in such an synergistic way  as to create a 
cyber-physical world \cite{IMT2030whitepaper}. Realizing the vision will require ubiquitous space-air-ground-sea coverage, very low latency, extremely  broad terahertz bands, and AI-native network architecture~\cite{YouXH2021Towards6G}. Furthermore, it  calls for seamless integration of communications, sensing, control, and computing. As a result, SemCom has the potential to play a pivotal role in 6G. The realization of SemCom will power several new types of 6G services, aiming at creating truly immersive experiences for humans, such as \emph{extended reality} (XR), high-fidelity holographic communications, and all-sense communications~\cite{ITU-T2019}. In the sequel, we will discuss the 6G services, their requirements for SemCom, and how they can be met by the  development of  6G core technologies. 

6G will feature a broad set of exciting new services and applications that extend  human senses in a fusion of the virtual and physical worlds. They include ubiquitous wireless intelligence, data teleportation, immersive XR, digital replication, holographic communications, telepresence, wearable networks, and sustainable cities. Several of them that are closely related to SemCom are described as follows, along with the new challenges they pose to SemCom. 

\begin{enumerate}
    \item \emph{Immersive XR:} XR is an umbrella term encompassing VR, AR, the \emph{mixed reality} (MR), and the intersections between them~\cite{Samsung20206GVision}. Boundless XR technologies will be integrated with networking, cloud/edge computing, and AI to offer truly immersive experiences for humans, applicable in a wide range of areas such as industrial production, entertainment, education, and healthcare. Its implementation requires the collection and processing of  data reflects or describes human movements and surroundings to generate  key features that guide  system operations, e.g.,  shifting rendered targets and displaying particular videos. Smooth human experience relies on  intentions and preferences that are being interpreted properly by devices and machines, so that they can produce and display desired contents. The continuous human-machine interaction places XR in the domain of H2M SemCom discussed in Section~\ref{Section:H2MSemCom}. Relevant designs  are similar to SemCom for VR/AR discussed therein but their requirements are more stringent in terms of accuracy and diversity  of sensing human characteristics (e.g. head movement, arm swing, gestures, speeches), data rates (e.g., 1Gbps for 16K VR~\cite{elbamby2018toward,Samsung20206GVision}) and latency (e.g., motion-to-photon delay below 15-30 ms~\cite{HuaweiXRlatency2018}). Moreover, the increased reliance of immersive XR on AI calls for SemCom design that is capable of a more efficient support of training and inference using large-scale AI models (i.e., scaling up SplitNet).

    \item \emph{High-fidelity Holographic Communication:} Holographic communication involves transmission of 3D holograms of human beings or physical objects. Based on high-resolution rendering, wearable displays, and AI, mobile devices will be able to render 3D holograms to display local presence of remote users or machines, creating a more realistic local presence of a remote human being or physical object~\cite{ITU-T2019}. Scenarios such as remote repair, remote surgey, and remote education can all benefit from this new form of communication~\cite{Tataria2021-6G}. This new form of SemCom aims at to enhancing visual perception of  users to improve the effectiveness of virtual interaction. This  requires high-resolution  encoding of  haptic information, colors, positions, and tilts of a human/object. Displaying  interactive high-fidelity holograms requires extremely high data rate (up to $4.3$ Tbps) and stringent latency constraints (possibly  sub-milliseconds)~\cite{Tataria2021-6G}. Such requirements make it crucial   to boost the efficiency  and speed of  semantic/effectiveness  encoding and transmission techniques  to unprecedented levels.  Moreover, since holographic communication can potentially involve both human users and machines, their real-time holographic interaction will require seamless integration of H2M and M2M SemCom techniques.

    \item \emph{All-Sense Communication:} 
    All of five senses, including sight, hearing, touch, smell, and taste will be included   in 6G communications using  an ensemble of  sensors that are wearable or mounted on each device. Combined with holographic communication, the all-sense information will be efficiently integrated to realize close-to-real feelings of remote environments \cite{IMT2030whitepaper ,Tataria2021-6G}. Such technologies will facilitate  \emph{tactile communications and haptic control}. In  all-sense communication,  the diversified  types of sensing signals create different new dimensions of information, resulting in exponential growth of the complexity of semantic information representation.

\end{enumerate}

The aforementioned future services presents formidable tasks for developing next-generation SemCom technologies. On the other hand, breakthroughs in the area are made possible  by leveraging the  revolution of  6G technologies. Some key aspects are described as follows. 

\begin{enumerate}
    \item \emph{Almost Limitless Connectivity:}  
    While 5G realizes  ubiquitous connectivity, 6G will strive to achieve \emph{almost limitless connectivity}. Specifically, in the 6G era,  we expect to experience enormously high bit rates of up to $1$ Tbps, low end-to-end latency  of less than 100 microseconds or high reliability with properly relaxed latency (e.g. $99.999\%$ with $3$ms in \emph{new radio vehicle}), high spectral efficiency of about $100$ bps/Hz, massive connections reaching at least $10^7\text{ devices/km}^2$, and ultra-wide and multi-frequency frequency bands of up to $3$ THz with air, space, earth, and sea coverage~\cite{6GSpeculative2020,YouXH2021Towards6G}. As a result, all machines and human beings will not only be connected  but do so in a profound, instantaneous  way as to enable in-depth knowledge sharing and interaction, large-scale collaboration, and extensive mutual care. Naturally, the advanced forms of SemCom techniques discussed in this article (for e.g., human-machine symbiosis and dialogues, human sensing and care, learning, inference, etc.) will benefit from almost-limitless connectivity and at the same time bring to it unprecedented end-to-end performance.

    \item \emph{Comprehensive AI:} AI has been established as a tool for solving problems originally intractable due to  either  prohibitive  complexity or the lack of models and algorithms. 6G  are being designed to be \emph{comprehensive AI systems} where  AI will  be extensively used for optimizing  the overall system performance and network operations \cite{Samsung20206GVision, 3GPP2019}. At the physical layer,  AI provides a data-driven approach for optimizing  modulation and channel coding. At the system level, AI models can automate the collaboration between devices and base stations. It is even possible to apply large-scale AI to optimize the end-to-end performance of a network by enabling, for example,  network self-recovering and self-organization. An AI comprehensive system, which  comprises a large number of wirelessly connected nodes/entities, intertwines machine learning, inference, and SemCom. For efficient implementation of such systems, it is essential  to have the availability of a rich library of advanced SemCom techniques  from which highly efficient effectiveness coding and transmission techniques can be retrieved and used  to support any of a wide-range of specific optimization tasks and network/system configurations with heterogeneous models and complexity. On the other hand, leveraging the omnipresence of AI, more complex and intelligent SemCom operations  can be realized to improve semantic and effectiveness encoding, thereby deepening  the level of H2H and H2M conversations, and narrowing the quality  gap between machine and human assistance and care. 

    \item \emph{Integrated Communication, Sensing, Control, and Computing:} 
    The realization the 6G applications (such as immersive XR and mobile holograms mentioned earlier) requires  resolving the conflict between the required  extensive computation capabilities and their reliance on many specialized low-cost, low-power edge devices. One  mainstream approach is to jointly design communication, sensing, control, and computing so as to improve the overall system performance under the devices' constraints. Another relevant approach is  to split  computing intensive  tasks and offload parts from devices to edge servers, which provide an edge computing platform,  for execution (which is aligned with the SplitNet approach discussed in this article). These approaches reflect the main theme  of the 6G innovation, namely the tight integration of different aspects of data processing and transportation. The required  deep application and semantic awareness by future wireless techniques will likely place SemCom at the central stage of 6G development.

\end{enumerate}

There is no doubt that SemCom will continue its growth, potentially becoming a primary area for technology innovation and breakthroughs in the 6G era. Coupling advanced SemCom and 6G technologies  paves the way towards the disappearance of the boundary between the physical and virtual worlds.


\bibliographystyle{IEEEtran}

\bibliography{SemCom-Submission-ArXiv}

\end{document}